\documentclass[11pt,preprint]{aastex}
\shortauthors{MCKEE, PARRAVANO, \&HOLLENBACH}
\shorttitle{Stars, Gas, and Dark Matter}

\newcommand{\beq}	{\begin{equation}}
\newcommand{\eeq}	{\end{equation}}
\newcommand{\beqa}{\begin{eqnarray}}
\newcommand{\eeqa}{\end{eqnarray}}
\newcommand{\avg}[1]  {{\langle #1 \rangle}} 

\newcommand{\e}	{$^{-1}$}
\newcommand{\ee}	{$^{-2}$}
\newcommand{\eee}	{$^{-3}$}
\newcommand{\tnm}{\tablenotemark}
\newcommand{\tnt}{\tablenotetext}

\newcommand{\caln}	{{\cal N}}
\newcommand{\calo}	{{\cal O}}
\newcommand{\calr}	{{\cal R}}
\newcommand{\pbyp}[1]	{{{\partial\hfil}\over{\partial#1}}}
\newcommand{\ppbyp}[2]	{{{\partial#1}\over{\partial#2}}}

\def\vecK{{\bf K}}

\def\vecnabla{
              \setbox1=\hbox{$\bigtriangledown$}
                           \raise.45ex\hbox{$\bigtriangledown$\hskip-.97\wd1
                           $\bigtriangledown$\hskip-.97\wd1
                           $\bigtriangledown$\hskip-.97\wd1}
                           \raise.47ex\hbox{$\bigtriangledown$}}
\def\div{{\vecnabla\cdot}}
\def\ltsimeq{\,\raise 0.3 ex\hbox{$ < $}\kern -0.75 em
 \lower 0.7 ex\hbox{$\sim$}\,}
\def\gtsimeq{\,\raise 0.3 ex\hbox{$ > $}\kern -0.75 em
 \lower 0.7 ex\hbox{$\sim$}\,}

\let\ga=\gtsimeq
\let\la=\ltsimeq
\def\sech{{\rm sech}}
\def\disk	{{\rm disk}}
\def\ddm	{{\rm DDM}}
\def\dm	{{\rm DM}}
\def\eff	{{{\rm eff}}}

\def\H	{{\rm H}}
\def\HI	{{\rm HI}}

\def\ind	{{\rm ind}}
\def\LB	{{\rm LB}}
\def\los	{{\rm los}}
\def\M	{{\rm M}}

\def\ms    {{\rm MS}}
\def\MS    {{\rm MS}}
\def\BD    {{\rm BD}}
\def\NS	{{\rm NS}}

\def\PMH11	{{\rm PMH11}}

\def\thin		{{\rm thin}}
\def\tot	{{\rm tot}}
\def\vis	{{\rm vis}}
\def\WD	{{\rm WD}}
\def\cnm	{{\rm CNM}}
\def\wnm	{{\rm WNM}}

\def\htwo	{H$_2$}
\def\juric	{Juri\'c\ }
\def\mbd	{{m_{\rm BD}}}
\def\fmM	{F_{m,\,\M}}

\def\mmk {{m_{\rm M-K}}}
\def\mpc	{\,M_\odot$~pc\ee}
\def\mpce	{\,M_\odot$~pc\eee}

\def\msun{$M_\odot \,$}

\def\nbh	{{\bar n_\H}}

\def\nhp	{N_{\H,\perp}}
\def\nhpt	{N_{\H,\perp}^{\rm thin}}

\def\nnso	{n_{*,0}}
\def\nnsm	{n_{*,{\rm M}}}

\def\nnsmo{n_{*,{\rm M},0}}
\def\rhosmo{\rho_{*,{\rm M},0}}

\def\dsst	{\dot N _{*T}}
\def\dSs	{\dot \Sigma _{*}}

\def\dssh	{\dot N _{*h}}
\def\ssm	{N _{*}(m)}
\def\dssm	{\dot N _{*}(m)}
\def\opo	{{1.1}}
\def\SSopo {\Sigma_{*1.1}}
\def\SSM {\Sigma_{*,\rm M}}

\def\ssPMH11	{N _{*,{\rm PMH11}}}
\def\ssoPMH11	{N _{*1,{\rm PMH11}}}
\def\mpre		{m_{\rm pre}}
\def\mpreo		{{m_{\rm pre,0}}}
\def\tpre		{t_{\rm pre}}
\def\taupre	{\tau_{\rm pre}}

\def\zmax{{z_{\rm max}}}
\def\rpc	{{\rm pc}}

\begin{document}

\title{Stars, Gas, and Dark Matter in the Solar Neighborhood}

\author{Christopher F. McKee\altaffilmark{1}, 
Antonio Parravano\altaffilmark{2}, and
David J. Hollenbach\altaffilmark{3}}

 \altaffiltext{1}{Physics Department and Astronomy Department
University of California at Berkeley, Berkeley, CA 94720} 

\altaffiltext{2}{Universidad de Los Andes, Centro De F\'{\i}sica
Fundamental, M\'erida 5101a, Venezuela}

\altaffiltext{3}{SETI Institute, 189 Bernardo Avenue, Mountain View,
 CA 94043}

\begin{abstract}

The surface density and vertical distribution of stars, stellar remnants, and gas in the solar vicinity 
form important ingredients for 
understanding the star formation history of the Galaxy as well as for inferring the local density of dark matter by
using stellar kinematics to probe the gravitational potential.   In this paper we review the literature for these baryonic components, reanalyze  data, and provide tables of the surface densities and exponential scale heights of main sequence stars, giants,
brown dwarfs, and stellar remnants.  We also review three components of gas (H$_2$, HI, and HII),
give their surface densities at the solar circle, and discuss their vertical distribution.
We find a local total surface density of M dwarfs of $17.3\pm 2.3$ M$_\odot$ pc$^{-2}$, significantly higher than previous values. Our result for the total local surface density of visible stars (main sequence stars and giants), 
$27.0 \pm 2.7$ M$_\odot$~pc\ee, is close to previous estimates due to a cancellation of
opposing effects: more mass in M dwarfs, less mass in the others.
The total local surface density in white dwarfs is 
$4.9\pm 0.6$~\msun~pc\ee; in brown dwarfs, it is $\sim1.2$ \msun~pc\ee,
but with considerable uncertainty.  
We find that the total local surface density of stars and stellar remnants is $ 33.4\pm 3\ M_\odot$~pc\ee,
somewhat less than previous estimates but within the errors of many of them.
We analyze data on 21 cm emission and absorption and obtain good agreement with recent results on the local
amount of neutral atomic hydrogen obtained with the {\it Planck} satellite.
The local surface density of gas is $13.7\pm1.6$ M$_\odot$~pc\ee.
The total baryonic mass surface density that we derive for the solar neighborhood is  $ 47.1\pm 3.4$ M$_\odot$ pc$^{-2}$ (43.8
 M$_\odot$ pc$^{-2}$ within 1.1 kpc of the midplane).
Combining these results with others' measurements of the total surface density of matter within 1-1.1 kpc of the plane, we find
that the local density of dark matter is $\rho_\dm=0.013\pm0.003\mpce\ $= 0.49\pm0.13$~GeV cm\eee.
The local density of all matter is
$0.097\pm0.013\mpce. 
We discuss limitations on the properties of a possible thin disk of dark matter.

\end{abstract}


\newpage

\section{Introduction}

A central goal in the study of the origin and structure of galaxies is to understand how both
baryonic and dark matter are distributed within them. For the Milky Way, essential input  is provided by the volume
density and surface density of stars, stellar remnants and gas in the solar vicinity---the local baryon budget. This input is
required for accurately inferring the amount of dark matter in the solar vicinity by dynamical modeling of observations of
stars in the gravitational
potential of both the dark matter and the baryonic matter, whether this
modeling is local, based on fitting observations of stars in the solar vicinity, or global, fitting the rotation
curve of the Galaxy (Read 2014).   
The value of $\rho_{{\rm DM}}$ is important 
for estimating the sensitivity
requirements of dark matter searches
and for constraining the shape of the Milky Way dark matter halo, which can be affected by late major mergers
(Read 2014).
It has also been suggested that there is a thin disk (scale height $h=\calo(100$ pc)--Shaviv et al. 2014) or very thin
disk ($h=\calo(10$ pc)--Fan et al. 2013; Randall \& Reece 2014; Kramer \& Randall 2015) of dark matter that brings the
period of vertical oscillations of the Sun into agreement with the $\sim 30$ Myr period of different phenomena observed on Earth.

The distribution of stars by mass (the present day mass function, or PDMF) 
provides important information on the star formation history of the Galaxy,
and the distribution of stars by distance from the Galactic plane provides information on the dynamical evolution of the stellar disk.
The shape of the PDMF also provides constraints on the stellar initial mass function (IMF). In this work, we shall assume that the IMF 
of stars is universal  in galactic disks when averaged over large regions of space
and long intervals of time (Bastian et al. 2010). Although significant variations in the stellar mass function of star
clusters are observed (Dib 2014; see also Parravano, McKee, \& Hollenbach 2011, hereafter PMH11), 
the summed mass distribution of objects from various clusters
converge to a smooth shape, the universal IMF.   
The stars in the solar neighborhood have been formed in
a huge number of events and environments during the lifetime of the local disk (e.g., Bland-Hawthorn et al, 2010), so
cluster-to-cluster variations should not affect our analysis.
No general trends of the shape of the mass function of stars in clusters
on the size of the cluster or its metallicity have been clearly established.   
The only deviations from a universal IMF that have been established are in massive elliptical galaxies
(e.g., Conroy et al. 2013, but see Smith et al. 2015)  
and near the Galactic Center (e.g., Lu et al. 2013), although it should be noted that there is no evidence for
deviations from a universal IMF in the old nuclear star cluster at the Galactic Center (Fritz et al. 2014).

A standard reference for the baryon budget in the solar neighborhood (which we define as a cylinder of radius 1 kpc
extending above and below the plane) is Flynn et al. (2006; hereafter F06). We make three improvements on their results:
First, we alter their PDMF so that it is consistent with the observations of stars
within 25 pc of the Sun as determined by Reid et al. (2002; hereafter RGH02). This involves both increasing
the surface density of M dwarfs based on a revision of
the analysis of Zheng et al. (2001; hereafter Z01) and decreasing the surface densities of more massive stars. 
Second, we determine the surface density of white dwarfs in two ways:
We implement the correction suggested by Katz et al (2014), who pointed out that a significant number of faint white
dwarfs in binary systems are missing from surveys, and then we calculate the number of white dwarfs expected given
the observed number of M dwarfs and a mildly decreasing star formation history; these two numbers are in good agreement.
Finally, we re-assess the distribution of gas in the solar neighborhood. We estimate the amount of HI that is missed due to
optical depth in the 21 cm line from two independent measurements, and we
are able to significantly reduce the uncertainty in
the column density of HI. 
We also give improved results for the vertical distributions of the gas and stars.

Beginning with Kapteyn (1922), astronomers have attempted to infer the density of dark matter in the solar vicinity, $\rho_\dm$, by analyzing
star counts and stellar kinematics. In the modern era, this study was rejuvenated by Kuijken \& Gilmore (1989a, 1989b, 1991),
who analyzed a sample of K dwarfs towards the south Galactic pole. Since then, a variety of techniques have been developed
and much larger samples of stars have been analyzed, as discussed in the comprehensive review by Read (2014). 
In many cases, analysis of stellar kinematics provides a value of the vertical gravitational acceleration, $K_z$. For a 
disk of constant surface density with a flat rotation curve,
this is equivalent to a measurement of the total surface density, $\Sigma$, since $K_z=-2\pi G\Sigma$. The 
surface density of the Galactic disk varies,
however, so this relation does not hold exactly (Kuijken \& Gilmore 1989a), and we evaluate the deviation.
As Read (2014) points out,
one obtains a better estimate of $\rho_{{\dm}}$ if one has a prior direct determination of the vertical density distribution of baryons, $\rho_b(z)$.
Our determination of the baryonic density distribution
can therefore tighten the constraints on $\rho_{{\dm}}$.  As a first step in that direction, we take some existing measurements of
the total surface density within 1 or 1.1 kpc of the plane and use our result for the baryonic surface density, $\Sigma_b$, to infer revised
values for $\rho_\dm$.

This paper is organized as follows.
We focus in Section 2 on the observed surface density of M dwarfs, which are the single largest contributor to the 
stellar surface density. 
Section 3 presents our determination of the surface density of white dwarfs.    
In Section 4
we summarize the surface densities of stars and stellar remnants, while Section 5 discusses the local interstellar gas.
We bring this all together in Section 6, where we present our results for the local baryon budget. In Section 7, we
apply these results to determining the local density of dark matter, and 
discuss the implications for disk dark matter, including the possible existence of a thin disk of dark matter in the Galaxy.
Our results are summarized in Section 8.
Appendix A discusses the variation of metallicity with height and shows that this has only a minor effect on the inferred
scale heights. In Appendix B we discuss the stellar halo and show that halo stars within about 3 kpc of the plane
are included in our analysis. 
Finally, in Appendix \ref{app:kz} we determine the relation between the vertical acceleration due to gravity and the surface density
of matter.

\section{The Surface Density of M Dwarfs}
\label{sec:M}

\subsection{Counting the M Dwarfs}
\label{sec:counting}

M dwarfs are the largest single contributor to the stellar mass, so we begin by estimating the surface
density of such stars in the solar neighborhood. We denote stellar masses, measured in solar masses, by $m$.
The mass range of M dwarfs extends from the maximum mass of a brown dwarf, $\mbd=0.075$ (Burrows et al. 2001),
which is almost independent of metallicity, to the minimum mass of a K dwarf, $\mmk$.
Following Gould et al (1996) and Zheng et al. (2001), we shall identify M dwarfs as those with 
absolute visual magnitudes $M_V\geq 8$. For the mass-luminosity relation adopted by Reid et al
(2002, hereafter RGH02), this corresponds to $\mmk=0.69$; for the mass-luminosity relation of Kroupa et al. (1993),
this corresponds to $\mmk=0.65$. 
We shall take the average of these and adopt $m_{\rm M-K}=0.67$.
This value applies in the solar neighborhood; the metallicity decreases away from the Galactic plane,
and so does the value of $\mmk$ (Appendix \ref{app:scale}).

Gould et al. (1996) used HST observations of distant field M dwarfs to make the first determination
of the surface density of M dwarfs in the solar neighborhood.
They inferred $M_V$ from a color-magnitude relation that was independent of height. Binaries were
not resolved in their observations, so they determined
``effective masses", that is, stellar masses  derived by assuming that binaries are single stars with the luminosity of the binary. They showed that the stellar mass surface densities inferred from the ``effective mass function" agree reasonably
well with the true surface densities. They used the Henry \& McCarthy (1993) mass-luminosity relation, for which
the maximum mass of an M dwarf (i.e., the dividing line between M and K stars) is
$\mmk=0.66$, which is very close to the value we have adopted for local stars, $\mmk=0.67$.
They found that simple one-component models for the number density of M dwarfs, $\nnsm$,
as a function of distance above the Galactic plane, $z$, such as 
$\nnsm\propto\sech^2(z/h)$ (expected for an isothermal, self-gravitating disk) or
$\nnsm\propto\exp(-z/h)$ (expected for isothermal stars in a constant gravitational field), are inconsistent
with the observations. 
(Throughout this paper we take $z\geq 0$ and assume that the matter distribution in the Galaxy is symmetric about the Galactic plane.)
However, simple two-component models, such as a double exponential or the sum of
a sech$^2$ term and an exponential,
\beq
\nnsm=\nnsmo\left[(1-\beta)\sech^2\left(\frac{z}{h_1}\right)+\beta\exp\left(-\frac{z}{h_2}\right)\right],
\label{eq:nnsm}
\eeq
are consistent with the data. 
Here 
$\nnsmo$ is the density of M dwarfs in the midplane,
$\beta$ is the fraction of the midplane density associated with the thick disk, and $h_2$ is taken to be greater than $h_1$.
They emphasized that the two terms do not necessarily correspond
to two physically distinct populations, and indeed Bovy, Rix, \& Hogg (2012b) have shown that there
is no distinct thick disk; instead, populations defined on the basis of metallicity have
a smooth distribution of scale heights.
Nonetheless, for simplicity we shall refer to these two terms as describing the thin disk and thick disk,
respectively.
The effective scale height for this number-density distribution (or for the sum of two
exponentials) is
\beq
h_\M\equiv\frac{N_{*,\rm M}}{2\nnsmo}=(1-\beta)h_1+\beta h_2\, ,
\label{eq:hsig}
\eeq
where $N_{*,\rm M}$ is the number of M dwarfs per unit area in the Galactic disk.
Under their assumption that
there is no metallicity gradient with height, the scale height $h_\M$ is the same
as the mass scale height,
\beq
h_{m,\M}=\SSM/(2\rhosmo).
\label{eq:hmm}
\eeq
For a distribution of the form in Equation (\ref{eq:nnsm}), they inferred that the local surface density of M dwarfs is
 $\Sigma_{*,\rm M}=12.4\pm 1.9\, M_\odot$ pc$^{-2}$, corresponding to an effective scale height
$h_{m,\M}=390$ pc. For
the double-exponential fit, they inferred a somewhat larger surface density, 
$\Sigma_{*,\rm M}=14.5\pm 2.4\, M_\odot$ pc$^{-2}$, but a considerably smaller effective scale height,
$h_{m,\M}=236$ pc.

As pointed out by Kuijken \& Gilmore (1989b), Gould et al. (1996), and Z01, stellar surveys are strongly
biased towards stars at large distances from the Galactic plane. 
Since the functional form of the density distribution in the disk is unknown, this
leads to a fundamental problem: Even with an accurate determination of the
stellar density away from the plane, it is difficult to accurately infer the density
near the midplane from such surveys, and correspondingly to accurately measure the
stellar surface density. Gould et al. (1996) used a clever trick to mitigate 
this problem: They fit the data to both a sech$^2$ + exponential (Equation (\ref{eq:nnsm}))
and to a double exponential.
Since these fits have different midplane densities, 
they were able to linearly combine them so as to match the observed midplane density.
It is important to note that although there are significant uncertainties in the distances to
the stars, and therefore in the stellar density, the uncertainty in the {\it number} of observed stars
is only statistical and is small. 
One can show that the number of observed stars is the same for the two fits
for the density,
and the linear combination preserves this number.
Gould et al. (1996) inferred that the contribution of the double exponential to the
total density was negligible by comparing their results
with the number density of early M dwarfs ($8.5\leq M_V\leq 12.5$) observed locally by 
Wielen, Jahreiss, \& Kr{\"u}ger (1983)--i.e., the sech$^2$--exponential
fit is in much better agreement with the data than the double exponential fit.
These results were revised slightly by Gould et al. (1997).

Z01 extended this work with more data. 
They considered two cases: one with
a constant color-magnitude relation, CMR(I), corresponding to no variation of metallicity with distance from
the plane, and one with a  $z$-dependent relation, CMR(II), which is more realistic 
(e.g., Bochanski et al. 2010; we use roman numerals to distinguish these CMRs in order
to avoid confusion with the notation for the thin and thick disks). 
For CMR(II), they also included an estimate of the $z$-dependence of
the mean mass-luminosity relation; a decrease in the metallicity results
in a lower mass for a given luminosity.
In order to eliminate contamination by spheroid stars, they considered only stars within a distance
$\zmax$ of the plane, with $\zmax=3200$~pc for CMR(I) and $\zmax=2400$~pc for CMR(II).
For CMR(I), they
found $\Sigma_{*,\rm M}=14.3\pm 1.3\, M_\odot$~pc\ee\ for the value of $\rhosmo=0.0180 M_\odot$~pc\eee\ inferred from Wielen et al. (1983), corresponding to $h_{m,\M}=397$~pc. 
For the $z$-dependent case, they used the $z$-dependent CMR(II) to infer the luminosities
and distances of the stars. They adopted a $z$-dependent mass-luminosity relation, but
rather than using that for each star, they applied it with the mean observed height in each
magnitude bin. The lower masses associated with lower metallicity led to a reduced
midplane density, $\rhosmo=0.0153 M_\odot$~pc\eee.  In this case
they found $\Sigma_{*,\rm M}=12.2\pm 1.6 M_\odot$~pc\ee, which
corresponds to an effective mass scale height $h_{m,\M}= 399$~pc, almost identical to that for CMR(I).

Subsequently, the properties of the stellar disk have been inferred from SDSS data.
\juric et al. (2008) inferred the properties of the stellar disk from observations of M1-M3 stars in
a specified range of $r-i$ colors, corresponding to a range of absolute magnitudes that was independent
of distance from the plane. 
Using a double-exponential fit, they
found $h_1=300\pm 60$~pc, $h_2=900\pm180$~pc, and $\beta=0.107$,
corresponding to $h_\M=364$~pc. In an independent analysis of SDSS data, Bochanski et al (2010) 
also used a double-exponential fit and found $h_1=300\pm15$~pc, $h_2=2100\pm 700$~pc, and $\beta=0.04$. Even though these properties of the thick disk are very different from those found by \juric et al. (2008), the effective scale height is very nearly the same: $h_\M=372$~pc. 
As noted by Bovy \& Rix (2013), the mass-weighted scale height of all the mono-abundance populations of Bovy et al. (2012b,c) is 400 pc.
Furthermore, Bovy \&Rix (2013) inferred a scale height of $370\pm 60$~pc
All these values are in 
good agreement with the results of Gould et al. (1996) and Z01.

\subsection{Revising the Results of Zheng et al. (2001)}
\label{sec:revise}

The measurement of the surface density of M dwarfs by Z01 can be improved
since they used pre-{\it Hipparcos} data from Wielen et al. (1983).
Jahreiss \& Wielen (1997) and RGH02 determined the local stellar luminosity function 
with the aid of {\it Hipparcos} data, although it should be noted that most of the
data on M dwarfs still comes from ground-based data.\footnote{Note that the more recent data imply
that the mass density of stars with $8.5\leq M_V\leq 12.5$, which Z01 used to normalize to the observations, is smaller than found by Wielen et al. (1983); however, the total
mass density of M dwarfs and their companions (i.e., excluding M-dwarf companions of
more massive stars) adopted by Z01 from Wielen et al. (1983) is very
nearly the same as that from Jahreiss \& Wielen (1997).}
Using the mass-luminosity relation proposed by RGH02, which is based on the results of
Delfosse et al. (2000) for $M_V>10$ and of Andersen (1991) for $M_V<10$, we find
that the local mass density of M dwarfs is 
 \beq
\rhosmo=0.0216~~~ M_\odot~\mbox{pc\eee} 
\eeq
from the RGH02 data (the earlier data of Jarhreiss \& Wielen (1997) gives
$0.0198 M_\odot$~pc\eee).
This density includes the contribution of the M-dwarf companions of stars more massive
than M dwarfs, which Z01 did not include; the data of RGH02 imply that this contribution is
$0.0012 M_\odot$~pc\eee, so that the density of M dwarfs excluding this contribution is about
$0.0204 M_\odot$~pc\eee, somewhat greater than the value
of $0.0180 M_\odot$~pc\eee\ adopted by Z01 for CMR(I).

We could follow a procedure similar to that of Gould et al. (1996) and Z01
to determine the M-dwarf surface density: combine the
sech$^2$--exponential and double-exponential fits so as to agree with the local density
of stellar systems observed by RGH02, and then use the mean mass per stellar system.
However, the much more straightforward approach of simply using the effective scale height
determined by Z01 and replacing the local
density that they used with the value measured by RGH02 gives very nearly the same answer,
so we shall use that. For both CMRs, Z01 found a scale height
$h_{\M}\simeq 400$~pc.
This scale height is somewhat larger
than the M-dwarf scale heights found by \juric et al. (2008) and Bochanski et al. (2010), but
this is consistent with the finding by Gould et al. (1996) and by Z01 that a sech$^2$--exponential fit gives a larger
effective scale height than the double exponential fit used by the former authors.
A virtue of the fit we have adopted from Z01 is that
it reduces the magnitude of the density
gradient at the midplane,
which is expected to vanish: 
For example, for the Z01 CMR(II) normalized density fit (i.e., the fit combining a sech$^2$-exponential
and a double exponential),
the relative gradient at the midplane, $(d/dz)(\nnsm/\nnsmo)$,
is 5 times smaller than for the \juric et al. (2008) or
Bochanski et al. (2010) double exponentials.

In Appendix \ref{app:scale} we show that the mass-weighted scale height,
$h_{m,\M}$, that is needed to infer the surface density is very close to $h_\M$.
Given the uncertainties, we shall adopt $h_{m,\M}\simeq h_\M\simeq 400$~pc based on
the results of Z01.
With this value of the effective scale height
and the local mass density of M dwarfs determined by RGH02, we conclude that the surface density of
M dwarfs is
\beq
\SSM=2\rhosmo h_{m,\M}=17.3 \pm 2.3~~~M_\odot\mbox{ pc\ee},
\eeq
where we have estimated the error based on the Z01 result for the metallicity-dependent CMR(II).
This value of the surface density is significantly higher than the value  14.3 $M_\odot\mbox{ pc\ee}$ of Z01 with CMR(I) and higher still than the value $12.2\, M_\odot\mbox{ pc\ee}$ of Z01 with CMR(II)
because it has a higher
midplane M-dwarf density and it includes the
$0.9 M_\odot$ pc\ee\ associated with the M-dwarf companions of more massive stars.

Our inferred value for the surface density of M dwarfs is the same for the two CMRs since
we take the midplane density from
observation and since the scale heights obtained by
Z01 are almost exactly the same for the two CMRs.
Why is it that Z01 found that the metallicity gradient significantly reduced
the inferred surface density, from $14.3\, M_\odot$~pc\ee\ to $12.2\,M_\odot$~pc\ee?
The primary reason for this is that they 
used the average {\it observed} height of the stars to infer the effect of the
metallicity gradient, whereas we use
the average  {\it actual} height. As they point out, their observations are biased toward
stars far from the plane; indeed, Figure 2 in Z01 shows that the average observed height
of the stars in their sample is about 1100 pc. On the other hand, the actual value of the average height 
of the stars is
of order the scale height, 400 pc, which is much less, thereby reducing the effect of the metallicity
gradient. A second difference between our analysis and theirs is that they found that
the metallicity variation had a substantial effect on the mass function, whereas our
assumption of a universal IMF means that there is no effect. As discussed in Appendix
\ref{app:scale}, in our work the effect of the metallicity gradient enters only in its reduction of
the maximum mass of an M dwarf
$\mmk(z)$ with height. The assumption that the IMF is independent of metallicity is supported
by the analysis of the mass functions of globular clusters, which have a wide range of metallicity
but have mass functions that depend primarily on the ratio of the time for the cluster to dissolve by
two-body encounters to the age of the cluster (de Marchi, Paresce, \& Portegies Zwart 2010),
not on the metallicity.

We are not aware of other direct determinations of the M-dwarf surface density that we can compare
to our result. We shall compare our result for the total stellar surface density with other results in Section
\ref{sec:comp}.

\section{White dwarfs} 
\label{sec:wd}

White dwarfs make a significant, but uncertain, contribution to the surface density. 
F06 combined white dwarfs, neutron stars and black holes into a single category, 
but as we shall see below,
this category is completely dominated by the white dwarfs.
We infer that they found
a total surface density for these objects of $6.9\, M_\odot$~pc\ee, which is 
is larger than the values given in the references below.
For white dwarfs, Sion et al. (2009) 
find a local density of $4.9 \times 10^{-3}$ WD~pc\eee, which is in good agreement 
with the previous estimate of $(5 \pm 0.7) 
\times 10^{-3}$ WD~pc\eee\ by Holberg et al. (2002),
although it is somewhat less than the estimate of $6.0\times 10^{-3}$ WD~pc\eee\ by Reid (2005).
Assuming a scale height  $h_1=325$ pc, Sion et al (2009) estimated that the surface density of white dwarfs is $N_{*,\WD}=3.2$  WD~pc\ee. They argued that there is no thick-disk contribution for white dwarfs, but this is difficult to accept in view of the well-established thick disk for the MS stars that are the precursors of white dwarfs (\juric et al. 2008, Bochanski et al. 2010). We note also that the vertical velocity dispersion 
of the DC and DQ white dwarfs in their sample exceeds 37~km~s\e, which corresponds to a scale height
significantly greater than 325 pc. Based on observations that extend well beyond the
20 pc local sample, a number of authors have presented
 direct evidence for a thick-disk component, with $\beta=0.2$ (Reid 2005), $\beta=0.16$ (Rowell \& Hambly 2011),
 and $\beta=0.33$ (Fuhrmann et al. 2012). 
If we combine the anomalously large halo component found by Rowell \& Hambly with their thick-disk component,
their value of $\beta$ would become 0.21.
 It is reasonable that $\beta$ is larger for white dwarfs than it is for main sequence stars since 
 on average the precursors to white dwarfs are older than main sequence stars. Since the white dwarfs observed in all these cases are almost all within 300 pc, it is not possible
 to infer the scale heights directly from the observations. The scale heights determined
 by \juric et al. (2008) give $h=420$~pc for $\beta=0.2$ and $h=500$~pc for $\beta=0.33$.
 In Section \ref{sec:age}, we infer that the scale height of white dwarfs is about 430 pc, close to
 the \juric et al. value for $\beta=0.2$. For the Sion et al. (2009) density, 
an effective scale height of 430 pc gives a surface density by number of 
$N_{*,\WD}=4.2$ WD pc\ee. 

Recently, Katz et al. (2014) have suggested that the number of white dwarfs could be underestimated by up to a factor 2 based on the difficulty in detecting white dwarfs in binary systems. They point out that the number of
bright white dwarfs with main sequence companions is about equal to the number of bright single white dwarfs, but the number of faint white dwarfs with such companions is much less than the number of faint singles. They argue that the number of white dwarfs in binaries is likely to be at least as large as the number of single white dwarfs, which leads to their estimate that the true number of white dwarfs has been underestimated by about a factor of two. A simple way of expressing their result is that if $n_{*,\WD,s}$ is the density of single white dwarfs and $f_{\WD,s}$ is the fraction of white dwarfs that are single, then the total density of white dwarfs is
\beq
n_{*,\WD}=\frac{n_{*,\WD,s}}{f_{\WD,s}}.
\eeq
Sion et al. (2009) find 89 isolated white dwarfs within 20 pc and argue that this is 80\% complete, which
implies that locally 
the density of single white dwarfs is
$n_{*,\WD,s}=3.3\times 10^{-3}$~WD~pc\eee. Katz et al. (2014) infer $f_{\WD,s}=0.4$ from Raghavan et al. (2010), which implies 
that the total density of white dwarfs is
$n_{*,\WD}=8.3\times 10^{-3}$~WD~pc\eee, 
which is 1.7 times larger than the
value of $4.9\times 10^{-3}$ WD pc\eee\
cited by Sion et al. (2009). For a white-dwarf scale height of $430$ pc, this corresponds to a surface
density of $N_{*,\WD}=7.1$~WD~pc\ee.

In view of the substantial discrepancy between the white-dwarf density found by Sion et al (2009) and that 
we infer from Katz et al. (2014), 
we now provide an alternative estimate of the white-dwarf density that is based on the IMF
and is independent of binarity.

\subsection{Ratio of White Dwarfs to M dwarfs}

We begin by calculating the surface density of white dwarfs produced by the stars that have formed
in the disk. We describe the IMF by $\psi(m)$, where $\psi(m)d\ln m$ is the fraction of stars (including
brown dwarfs) born between $m$ and $m+dm$ and where $m$ is measured in solar masses.
The rate at which stars are born per unit area in the mass range $dm$ is then
\beq
d\dssm=\dsst\psi(m)d\ln m,
\label{eq:imf}
\eeq
where $\dsst$ is the total rate at which stars are born per unit area. The star formation rate (SFR) 
per unit area in solar masses per
unit time is $\dSs=\dsst\avg{m}$, where $\avg{m}\sim 0.5$ is the average mass in the IMF.

Most IMFs (e.g., Kroupa (2002), Chabrier (2005), and PMH11)\footnote{The Kroupa (2002) IMF is a
broken power law, $\psi\propto m^{\gamma_i}$, with 4 segments; the Chabrier (2005) IMF is
a log normal function of $m$ for $m<1$ and a power law for $m>1$; and the PMH11 IMF has
the form $\psi\propto m^{-\Gamma}[1-\exp-(m/m_{\rm ch})^{\gamma+\Gamma}]$, which smoothly
transitions from $\psi\propto m^\gamma$ at low mass to $\psi\propto m^{-\Gamma}$ at high mass.
We have changed notation from PMH11, replacing $\varsigma_*(m)$ with $dN_*(m)/d\ln m$.}
 have a power-law form,
$\psi\propto m^{-\Gamma}$, for $m_u>m\ga 1$, where $\Gamma=1.7$ for the Kroupa (2002) individual star
IMF and $\Gamma=1.35$ for the other two IMFs;
$m_u$, the upper bound on the stellar mass in the power-law regime of the IMF, exceeds $100\, M_\odot$.
In this power-law regime, the stellar birthrate per unit area can be
written as
\beq
d\dssm=B\dSs m^{-\Gamma}d\ln m,
\eeq
where $B$ is a constant that depends on the IMF. One way to characterize IMFs is through the total
mass of stars formed per high-mass star, $\mu_h$. We define a ``high-mass star" as one with $m_u\geq m\geq m_h$,
and choose $m_h=8$ so that the number of high-mass stars corresponds to the number of core-collapse
supernovae, neglecting the complications associated with binary evolution (Sana et al. 2011). The rate of high-mass
star formation per unit area is
\beq
\dssh=\int_{m_h}^{m_u} B\dSs m^{-\Gamma}d\ln m=\frac{B\dSs\phi_h}{\Gamma m_h^\Gamma},
\eeq
where
\beq
\phi_h=1-\left(\frac{m_h}{m_u}\right)^\Gamma
\eeq
is close to unity. The mass of stars formed per high-mass star is 
$\mu_h\equiv\dot \Sigma_{*}/\dssh$, so that
\beq
B=\frac{\Gamma m_h^\Gamma}{\mu_h\phi_h}.
\eeq
The quantity $B$ is relatively constant for the three IMFs we are considering: $B=(0.276,\,0.255,\, 0.237)$ for
the Kroupa (2002), Chabrier (2005) and PMH11 individual star IMFs, respectively, whereas
$\mu_h$ varies by more than a factor 2: $\mu_h=(213,\,90, \,97)$.

We describe the star-formation history in terms of
the ratio of the star-formation rate to the value averaged over the age
of the disk at the solar circle, $t_0$ (cf. Miller \& Scalo 1979),
\beq
b(t)\equiv\frac{\dsst(t)}{\avg{\dsst}},
\label{eq:bt}
\eeq
where
\beq
\avg{\dsst}\equiv\frac{1}{t_0}\int_0^{t_0}\dsst(t)\;dt.
\eeq
Note that $t_0$ corresponds to the present time; it is the time that has elapsed since the onset of star formation in the solar vicinity. We further note
that by definition we have 
\beq 
\label{eq:def}
\frac{1}{t_0}\int_0^{t_0} b(t)\; dt =1. 
\eeq

The birthrate per unit area 
of white dwarfs at time $t$ due to stars which had a main-sequence mass 
in the mass range $\mpre$ to $\mpre+d \mpre$ is
\beq
d\dot N_\WD(\mpre,t) = \dsst(\tpre)\psi(\mpre)d\ln\mpre,
\eeq
where $\tpre=t-\taupre$ is the time at which the precursor was born and
$\taupre$ is the lifetime of the white-dwarf precursor of mass $\mpre$, 
including post main-sequence evolution.
The lowest-mass white dwarf precursor has a precursor lifetime equal to the age of the disk, $t_0$; label
this mass $\mpreo$. Under the assumption that white dwarf precursors have
a maximum mass of $8\,M_\odot$, the total surface density of white dwarfs is then
\beqa
N_\WD&=&B\avg{\dot\Sigma_*}\int_{\mpreo}^8 \mpre^{-\Gamma}\,d\ln\mpre \int_0^{t_0-\taupre} b(\tpre)d\tpre,\\
&\equiv&a_1 B\avg{\dot\Sigma_*} t_0,
\label{eq:a1}
\eeqa
where 
$\avg{\dot\Sigma_*}$ is the star-formation rate averaged over the time $t_0$ and
$a_1$ is a dimensionless
parameter of order $\mpreo^{-\Gamma}$, which as we shall see is of order unity. 
Adopting a power-law form for the precursor lifetime
as a function of mass, 
\beq
\taupre=\left(\frac{\mpreo}{\mpre}\right)^{\ell} t_0,
\eeq
we find that in the mass range $0.9<\mpre<1.6$, setting $\ell=3.5$ and
$\mpreo=(1.06,\,0.99)$ for metallicities of $Z=(0.019,\,0.008)$, respectively,
gives an accuracy better than $\sim$ 6\%
based on the evolutionary tracks of Girardi et al. (2000). 
Given the weak dependence of $\mpreo$ on metallicity, we shall neglect it and set $\mpreo=1$ in
our numerical evaluations.

As a simple model for a variation in the star formation rate, we adopt a linear form for $b(t)$ (Eq. \ref{eq:bt}),
\beq
b(t)=2-b_0-2(1-b_0)t/t_0,
\label{eq:boft}
\eeq
where $b_0=b(t_0)$ is the current value of $b(t)$.
Then the coefficient $a_1$ in Equation (\ref{eq:a1}) can be evaluated as
\beq
a_1\simeq\frac{\ell(2\ell+2\Gamma-\Gamma b_0)}{\Gamma(\Gamma+\ell)(\Gamma+2\ell)\mpreo^\Gamma}-\frac{1}{\Gamma 8^\Gamma},
\eeq
where we have ignored terms of order $8^{-\ell}$, and we have adopted $\ell\simeq 3.5$ for $1.6<\mpre<8$
since that mass range does not contribute much to the integral.
For a constant star formation rate
($b_0=1$) and $\mpreo\simeq 1$, this gives $a_1=0.49$, whereas in the opposite case in which
the current SFR is very small ($b_0\simeq 0$) we have $a_1=0.58$.
 
We now compare the predicted surface density of white dwarfs with the surface density of M dwarfs,
\beq
\SSM=\fmM\avg{\dSs}t_0,
\label{eq:FmM}
\eeq
where $\fmM$ is the M-star mass fraction produced by a given IMF.
In writing this equation, we have ignored the weak dependence of $\fmM$ on metallicity; the
effect of the variation of mean metallicity with height is discussed in Appendix \ref{app:scale}.
For the Kroupa (2002), Chabrier (2005) and PMH11 IMFs, $\fmM$  is given by
$\fmM=(0.45,\,0.29,\,0.33)$, respectively.
Inserting Equation (\ref{eq:FmM}) into
Equation (\ref{eq:a1}), we find that
the ratio of the surface density of white dwarfs by number to the surface density of M dwarfs by mass is
\beq
\frac{N_\WD}{\SSM}=\frac{a_1 B}{\fmM}.
\label{eq:sigwdm}
\eeq
Alternatively, in terms
of the midplane densities, we have
\beq
\frac{n_{\WD,0}}{\rho_{*,\M,0}}=\left(\frac{a_1 B}{\fmM}\right)\frac{h_{m,\M}}{h_\WD}.
\label{eq:nwdm}
\eeq
Although the white dwarf scale height, $h_\WD$, is not directly measured, its value relative to that for M
stars can be inferred, as we shall see below.

\subsection{The Mean Age and Scale Height of White Dwarfs}
\label{sec:age}

It remains to estimate the ratio of the scale heights of white dwarfs and M dwarfs. The velocity
dispersion of stars is observed to increase with age; from an analysis of the 16682 nearby F and G
dwarfs in the Geneva-Copenhagen Survey of the Solar Neighborhood (Nordstr\"om et al 2004),
Holmberg et al. (2009) found that the vertical velocity dispersion, $\sigma_W$, increases as
age$^{0.53}=(t_0-t_*)^{0.53}$, where $t_*$ is the time at which the star was born, so that
$t_0-t_*$ is its age. We find that the results of F06 imply that the scale height, $h$,
scales as $\sigma_W^{1.38}$ for stars with $M_V>3$, so that $h\propto (t_0-t_*)^{0.73}$. 
For a white dwarf, the relevant
age is that of its precursor, $t_0-\tpre$. Averaging over all the white dwarfs and M dwarfs, we have 
\beq
\frac{h_\WD}{h_\M}\simeq\left(\frac{\avg{t_0-\tpre}}{\avg{t_0-t_{*,\M}}}\right)^{0.73},
\label{eq:hfrac}
\eeq

For the form for $b(t)$ that drops linearly from $t=0$ to $t=t_0$ that
we have adopted (Eq. \ref{eq:boft}), the average birth time of M dwarfs is
\beq
\avg{t_{*,\M}}=\frac{1}{3}\left(1+\frac 12 b_0\right)t_0,
\eeq
which corresponds to an average age of $\frac 12 t_0$ for a constant SFR ($b_0=1$) 
and $\frac 23 t_0$ for $b_0=0$ (recall $b_0=b(t_0))$.
The average birth time of the stars that became white dwarfs is
\beqa
\avg{\tpre}&=&\frac{1}{N_\WD}\int_\mpreo^8 d\ln m \int_0^{t_0-\taupre} \dot N_*(m,\tpre) \tpre d\tpre,\\
&=&\frac{B\avg{\dot\Sigma_*(t)}}{N_\WD}\int_\mpreo^8\frac{d\ln \mpre}{\mpre^\Gamma}\int_0^{t_0-\taupre}b(\tpre)\tpre d\tpre,\\
&\equiv& \frac{a_2 B\avg{\dot\Sigma_*(t)} t_0^2}{N_\WD},\\
&=&\frac{a_2 t_0}{a_1},
\eeqa
where
\beq
a_2\simeq \frac{\ell^2}{\Gamma(\Gamma+2\ell)(\Gamma+3\ell)\mpreo^\Gamma}\left[2+\frac{(\ell-\Gamma)b_0}{\Gamma+\ell}\right]-\frac 13\left(1+\frac 12 b_0\right)\frac{1}{\Gamma 8^\Gamma},
\eeq
where again we have 
ignored terms of order $8^{-\ell}$ and have adopted $\ell\simeq 3.5$ for $1.6<\mpre<8$.
For a constant SFR and $\mpreo\simeq 1$, this corresponds to an average age 
of white-dwarf precursors of $0.59 t_0$, whereas for $b_0=0$, the
average age is $0.71t_0$.
The ratio of the age of the white-dwarf precursors to that of the M dwarfs is
\beq
\frac{\avg{t_0-\tpre}}{\avg{t_0-t_{*,\M}}}=\frac 32\frac{ a_1-a_2}{a_1(1-\frac 14 b_0)},
\label{eq:age}
\eeq
which is 1.18 for $b_0=1$ and 1.06 for $b_0=0$---i.e., $1.12\pm 0.06$ for $1\geq b_0\geq 0$.
Normalizing to the case $b_0=\frac 12$, which is 
consistent with the results of Aumer \& Binney (2009)
and approximately the value we get in Parravano, McKee \& Hollenbach (in preparation), we find that
Equations (\ref{eq:hfrac}) and (\ref{eq:age}) imply that 
for $\Gamma=1.35$
the ratio of the scale heights is 
\beq
\frac{h_\WD}{h_\M}\simeq 1.085-0.080\left(\frac 12-b_0\right)
\eeq
to within 0.1\%.
In Section \ref{sec:revise}, we adopt $h_\M=400$~pc for the M-dwarf scale height; for
$b_0=\frac 12$, this gives $h_\WD=434$~pc.

\subsection{Surface and Volume Densities of White Dwarfs}
\label{sec:surface}

In terms of the observed surface density of M dwarfs, the predicted surface density of
white dwarfs for the PMH11 IMF is (Eq. \ref{eq:sigwdm})
\beq
N_\WD=\left[6.7+1.1\left(\frac 12-b_0\right)\right]\left(\frac{\SSM}{17.3\,M_\odot\, \mbox{pc\ee}}\right)
~~~~~\mbox{WD pc\ee}.
\eeq
 The local volume density of white dwarfs is (Eq. \ref{eq:nwdm})
\beq
n_{\WD,0}=\left[7.6\times 10^{-3}+1.8\times 10^{-3}\left(\frac 12-b_0\right)\right]\left(\frac{\rhosmo}{0.0216\,M_\odot\,\mbox{pc\eee}}\right)~~~~~\mbox{WD pc\eee}.
\label{eq:nwdo}
\eeq
For the Chabrier (2005) individual-star IMF, $N_\WD$ and $n_\WD$ are increased by a factor 1.21.

We conclude that if the star formation rate has decreased linearly in time over the
life of the disk so that it is now half the average rate ($b_0=0.5$), then
the PMH11 IMF implies that the density of white dwarfs is $n_{*,\WD}=7.6\times 10^{-3}$~pc\eee,  
and the Chabrier (2005) individual-star IMF implies that it is $n_{*,\WD}=9.3\times 10^{-3}$~pc\eee. 
Equation (\ref{eq:nwdo}) shows that this density estimate is
not sensitive to the exact value of the current rate divided by the average rate.
We shall adopt the average of these two values, $n_{*,\WD}=8.5\times 10^{-3}$~pc\eee, which is very close to
the value of $8.3\times 10^{-3}$~WD pc\eee\ that we inferred from Katz et al. (2014). This analysis confirms their conclusion that
the number density of white dwarfs has been significantly underestimated by a number of workers.
To obtain a rough estimate of the error in these numbers we add in quadrature the uncertainty in the column density of M dwarfs (2.3/17.3) and
the uncertainty in the IMF, which we take to be 10\% based on the difference between our adopted density
and the values implied by either the PMH11 or Chabrier IMFs. The result is an uncertainty of 17\%.

To convert these results for number densities to mass densities,
we adopt the Holberg et al. (2008) value for the mean
mass of local white dwarfs, $0.665\, M_\odot$. 
This value is inferred from a catalog of white dwarfs within 20 pc of the Sun, which they infer to be 80\% complete;
this catalog is almost identical to the later catalog of Sion et al. (2009). The masses are determined from the stellar surface gravities
and effective temperatures; in addition, most of the stars have trigonometric parallaxes. The masses of the white
dwarfs with well determined masses (uncertainties less than 10\%) in this catalog range from $0.40\,M_\odot$ to $1.25\,M_\odot$.
Over half the stars have mass uncertainties less than $0.04\,M_\odot$.
With this value of the mean mass of white dwarfs, we find
that the local mass volume density is
$\rho_{*0,\WD}=0.0056\pm 0.0010\, M_\odot$~pc\eee\ and that the local 
mass surface density is $\Sigma_{*,\rm WD}=4.9\pm 0.8$~\msun~pc\ee.

\section{Stellar Surface Densities}
\label{sec:sigma}

The standard reference for stellar surface densities in the solar neighborhood is F06, 
which builds on the previous work by Holmberg \& Flynn (2000, 2004). These authors
assumed that each stellar component was isothermal and related the local mass density to the surface
density by solving the Poisson-Boltzmann equations.
For M dwarfs, they took the surface density, $\SSM$, from the work
of Gould et al. (1998) and then inferred the local mass density
of these stars, $\rhosmo$, whereas for all other stars they
inferred the surface density from the observed local density and the calculated
scale height. Holmberg \& Flynn (2004) introduced
a thick stellar disk in their model for the vertical distribution of stars. 
F06 updated this model, and from their results we infer that the thick disk has 10\% of the local volume
density of stars ($\beta=0.1$) and a scale height $h_2=1000$~pc; they assumed that 
main-sequence stars with
$M_V<4$ were too young to have a thick-disk component. 
F06 find $28.5\, M_\odot$~pc\ee\ in the thin disk and $7\, M_\odot$~pc\ee\ in the thick disk.
They do not specify the composition of the thick disk, although it is clear that it has no main sequence stars
with $M_V<4$. We assume that the other stars in the thick disk, which are all long-lived,
have the same proportions as in the thin disk. (Since the stars in the thick disk are older than those
in the thin disk, the proportion of red giants and white dwarfs in the thick disk should be 
somewhat larger, but we omit that complication here.) 
After distributing long-lived stars from the thick disk in proportion to their abundance in the thin disk,
we represent the F06 model as shown in Table \ref{tab:local}. 
F06 include halo stars as a separate category. As discussed in Appendix
\ref{app:halo}, determinations of the surface density of stars generally consider stars below $\sim$ 3 kpc;
if halo stars are not included as a separate category, they are included in the thick disk. About 2/3 of the
stars in F06's halo are within 3 kpc of the plane, and they are included in the table.
The value of the local surface
density of stars and stellar objects found by F06 is $36.1\,M_\odot$~pc\ee. 
It is of interest to compare their result for main sequence stars to that found many years ago by
Miller and Scalo (1979). These authors found a surface density of $27\, M_\odot$~pc\ee; since they
estimated the scale height of these stars as 325 pc, this does not include halo stars. The
F06 value for main-sequence stars
in the disk
(in Table \ref{tab:local}, visible stars minus giants and halo stars),
$25.8\, M_\odot$~pc\ee, is quite close to this.
As we shall see, our value for this quantity, $26.4\mpc, is even closer to the Miller \& Scalo (1979) value.

\begin{deluxetable}{lcccccc}
\tablecolumns{7}
\tablecaption{Local Stellar Surface Densities (Excluding Halo Stars with $z>3$ kpc)\tnm{a}
\label{tab:local}}
\tablehead{
\colhead{~~Description~~}&
\colhead{~~$\Sigma_*$(F06)\tablenotemark{{\rm b}}~~}&
\colhead{~\vline~~}&
\colhead{~~$\rho_{*0}$~~}&
\colhead{~~$h$~~}&
\colhead{~~$\Sigma_*$~~}&
\colhead{~~$\Sigma_{*1.1}$~~}\\
\colhead{}&
\colhead{($M_\odot$ \mbox{ pc\ee})}&
\colhead{\vline}&
\colhead{($M_\odot$ pc\eee)}&
\colhead{(pc)}&
\colhead{($M_\odot$ pc\ee)}&
\colhead{($M_\odot$ pc\ee)}
}
\startdata
$M_V<3$&1.5&\vline& 0.0018& 140& 0.5&0.5\\
$3<M_V<4$&1.1&\vline& 0.0018& 236& 0.8&0.8 \\
$4<M_V<5$&2.2&\vline& 0.0029&384&2.2&2.1\\
$5<M_V<8$&7.2&\vline& 0.0072&400&5.8& 5.4\\
$8<M_V$ (M dwarfs) &13.8&\vline&0.0216&400&17.3&16.2\\
Giants&0.5&\vline&0.0006&344&0.4&0.4\\
Halo ($z<3$ kpc)\tnm{c} &0.4&\vline &...&...&0&0\\
\hline
Visible stars &26.7& \vline &0.036&...&27.0&25.4\\
\hline
Brown dwarfs (BD)&2.3&\vline&0.0015&400&1.2&1.1\\
White dwarfs (WD)&6.9\tablenotemark{d}&\vline&0.0056&430& 4.9&4.5\\
Neutron stars (NS)&...&                \vline&  0.0001       &  ... &0.2&0.1\\
Black holes (BH)&...&                 \vline&       0.0001  &  400\tablenotemark{e} &0.1&0.1\\
\hline
Total&35.9&              \vline &    0.043         & ...   & $33.4\pm 3$    &    $31.2\pm 3$
\enddata
\tnt{a}{ $\Sigma_*$ is the local stellar surface density
based on F06. Numbers to the right of the vertical line are based on our analysis: 
$\rho_{*0}$ is the local stellar mass density in the midplane, $h$ is the effective scale height,
$h=\Sigma_*/(2\rho_{*0})$, and $\Sigma_{*,1.1}$ is the local surface density of stars within 1.1 kpc of the midplane.}
\tablenotetext{{\rm b}}{The $7 \,M_\odot$ pc\ee\ in the F06 thick disk has been distributed to the different categories as described in the text. The F06 halo stars above 3 kpc ($0.2\,M_\odot$~pc\ee) have been excluded.}
\tnt{c}{Halo stars within about 3 kpc of the Galactic plane are included in our entries for stars and stellar remnants (Appendix \ref{app:halo}).}
\tablenotetext{d}{This value includes neutron stars and black holes.}
\tablenotetext{e}{The scale height of black holes is unknown and has been assumed equal to that of M dwarfs.}
\end{deluxetable}

\subsection{Main-sequence stars and giants}
\label{sec:msg}

We update the F06 results in several ways. First, we use the local densities of main-sequence stars
measured by RGH02 (see Table \ref{tab:local}). 
The local density of main sequence stars we calculate from the RGH02 results is $\rho_{*0}({\ms})=0.035\, M_\odot$~pc\eee, which
is the same as that from F06;\footnote{RGH02 quote a lower value, however: $\rho_{*0}=0.031\,M_\odot$~pc\eee.} adding
$0.0006\, M_\odot$~pc\eee\ for giants then brings the total local density of visible stars up to $0.036\,M_\odot$~pc\eee.
The local density of M dwarfs measured by RGH02 is larger than that adopted by F06, and as a result 
the M-dwarf surface density we determined
in Section \ref{sec:revise}, $\SSM=17.3 \,M_\odot$~pc\ee, is larger than the value we infer from the F06 model,
$\SSM=13.8\, M_\odot$~pc\ee. 
The RGH02 density for stars with $M_V<3$ is significantly less than
that used by F06. However, 
in their Table 3, F06 cite a subsequent, more complete analysis of these
stars that is within 10\% of the RGH02 value, so we regard the value we adopt as more accurate.

Next, we reduce the F06 surface densities for stars earlier than M dwarfs by adjusting their scale
heights to be consistent with more recently determined values.
For low-mass stars with $M_V>4$, F06 found that the scale height of the thin disk was $h_1\simeq 400$~pc. 
However, as discussed
above, two subsequent analyses have found that in fact $h_1\simeq 300$~pc 
(\juric et al. 2008, Bochanski et al. 2010). Other than the larger value of the thin-disk scale height in
their model, it is very similar to the model of \juric et al. (2008): $\beta=0.1$ vs. $\beta=0.107$ and
$h_2=1000$~pc vs. $h_2=900$~pc. Although Bochanski et al. (2010) found a very different
model for the thick disk than \juric et al. (2008), the effective scale heights were almost the same:
$h=364$~pc for \juric et al. (2008) and $h=372$~pc for Bochanski et al. (2010).
By contrast, the effective
scale height for stars with $M_V>5$ implied by the F06 model is $h=(0.9\times 405+0.1\times 1000)$~pc $=464$~pc.
Previous authors had also found smaller scale heights than F06:
Recall that Z01
found $h\simeq 400$~pc;
in another case, RGH02 cite evidence that the scale height for stars with $M_V<3$ is only 100 pc,
whereas the F06 scale height for stars with $M_V<2.5$ is 145 pc.
A strength of the F06 analysis is that it is based on solving the Poisson-Boltzmann equation for
each type of star separately. We therefore assume that the {\it relative} values of the F06
scale heights are correct, and adjust the overall normalization. 
To be consistent with our analysis for M dwarfs, we adopt $h=400$ pc for the scale height of stars with $M_V>5$ 
and reduce the effective scale heights for stars
with $M_V<5$ by a factor $400/464=0.86$.  
The resulting scale heights are given in Table \ref{tab:local};
we have followed F06 in assuming that the thick disk has no main-sequence stars
with $M_V<4$. To obtain the surface
densities, we combine these revised scale heights with the stellar densities from RGH02.  The corrected scale height for stars with $M_V<3$, which is 140 pc, is significantly 
greater than the value cited by RGH02, so it is possible that the surface density of these stars is over-estimated.
For giant stars, we use the value of the local density given by F06; the reduction in the effective
scale height for these stars reduces their surface density from $0.5\, M_\odot$~pc\ee\ to
$0.4\, M_\odot$~pc\ee.
Overall, since our
value for the M-dwarf surface density is larger than the value F06 inferred from Gould et al. (1998),
our value for the surface density of main-sequence stars, $\Sigma_{*,\MS}=26.7 \, M_\odot$~pc\ee, is
slightly higher than their value, $\Sigma_{*,\MS}=25.8 \, M_\odot$~pc\ee.
The F06 model also includes a small contribution from the stellar halo; as discussed in Appendix \ref{app:halo},
halo stars within about 3 kpc of the Galactic plane are included in our results for stars and stellar remnants in Table \ref{tab:local}.

\subsection{Brown dwarfs.} 
\label{sec:bd}
The density of brown dwarfs is uncertain, since the fraction of stars that are brown dwarfs
is inferred to be smaller in the field than in young clusters. In PMH11, we determined that
observations of young clusters imply that the number ratio of low-mass
main-sequence stars ($0.08<m\leq1.0$) to stars (mainly brown dwarfs) in the mass range
$0.03<m<0.08$ is $R_\BD^\ind=3.76$, where the superscript
indicates that this number counts the individual stars in binaries. On the other hand, the ratio of
the number of main-sequence stars with $m\leq 1$ (almost all of which have $m>0.08$) to the
number of brown dwarfs in the Kirkpatrick et al. (2012) catalog of stars within 8 pc (almost all
of which are in the mass range $0.03<m<0.075$) is 5.8, corresponding to fewer brown dwarfs. In order to infer the mass density
of brown dwarfs,
it is also necessary to know the distribution of brown-dwarf masses, which 
is often assumed to be a power law, $d\caln_*/dm\propto m^{-\alpha}$,
where $\caln_*(m)$ is the number of stars with masses less than $m$.
In the limit of small masses, the IMF of PMH11 approaches the form $d\caln_*/d\ln m\propto m^\gamma$,
so that $\alpha= 1-\gamma$. In PMH11, we estimated $\gamma=0.51$, corresponding to
$\alpha=0.49$; this is in good agreement with the value $\alpha=0.51^{+0.24}_{-0.27}$ determined from
microlensing observations by Sumi et al. (2011). The power law inferred from the microlensing obsevations
is also consistent with the Kroupa (2002) power law in this mass range, $\gamma=0.7\pm 0.7$.
However, Kirkpatrick et al. (2012) infer 
$\alpha\sim -1$, corresponding to $\gamma\sim 2$.  The observational estimates for the density of brown
dwarfs are thus uncertain due to the apparent contradiction between the values obtained from young clusters and
microlensing on the one hand and
those obtained from infrared observations of the field on the other.
We adopt the value $\gamma=0.51$ given by
the PMH11 IMF, which is consistent with observations of brown dwarfs in young clusters, with the microlensing
observations, and with Kroupa's (2002) value.
For this choice,
the total mass surface density of brown dwarfs is about $\Sigma_{*,\BD}= 1.2$ \msun~pc\ee\ 
and the mass surface density at $z<1.1$ kpc is 1.1 \msun~pc\ee,
but with considerable uncertainty in both values.  
For example, if we ignored the result of Sumi et al. (2011) and adopted the result of Kirkpatrick et al. (2012) for the
density of brown dwarfs, then these values of the surface density would be reduced by a factor 3.76/5.8=0.65.
We obtained the mass surface density at $z<1.1$ kpc  by
assuming that the scale height of BDs is equal to that of M dwarfs, 400 pc, so that 93.6\% of the BDs are within
1.1 kpc of the mid plane.

\subsection{Neutron Stars and Black Holes}

Neutron stars are often born with high velocities, so it is not possible to directly relate the surface
density of neutron stars to that of their progenitors. Sartore et al. (2010) have considered a range of
models for the velocity distribution of neutron stars at birth and have determined the local density distribution
as a function of distance from the plane. We omit their case 1E, which has a scale height for neutron stars
of only 33 pc.   They assumed a  fiducial
case in which there are a total of $N_{\NS}=10^9$ neutron stars in the Galaxy, and their results scale
with $N_9= N_{NS}/10^9$.  Evaluating the surface density of neutron stars 
within 1.1 kpc of the plane in their models and assuming a mean neutron star
mass of $1.4 M_\odot$, we find
$\Sigma_{*1.1,\NS}=(0.3\pm 0.1)N_9 \, M_\odot$~pc\ee\  for the local mass surface density of neutron stars
within 1.1 kpc of the plane.  The average total surface density
in the Sartore et al. (2010) models we consider is twice this, $\Sigma_{*,\NS}=(0.6\pm 0.2)N_9 \, M_\odot$~pc\ee. 
Ofek (2009) performed simulations very similar to Sartore et al. (2010), and the average of his two
simulations give $\Sigma_{*1.1,\NS}=(0.55\pm 0.05)N_9 \, M_\odot$~pc\ee and
$\Sigma_{*,\NS}=(0.97\pm 0.13)N_9 \, M_\odot$~pc\ee.  Averaging the values from
Ofek and Sartore et al. (2010), we find $\Sigma_{*1.1,\NS}=(0.4\pm 0.11)N_9 \, M_\odot$~pc\ee
and $\Sigma_{*,\NS}=(0.8\pm 0.24)N_9 \, M_\odot$~pc\ee.   Similarly, we average the values
of Sartore et al. (2010) and Ofek to obtain the local midplane density of neutron stars.   We find
$\rho_{*0,\NS}= (4.7\pm1.6)\times 10^{-4}N_9 \, M_\odot$~pc$^{-3}$.    Using the PMH11 IMF and the Hurley
et al (2000) evolution code, which relates the mass of the progenitor to the likelihood
of producing a neutron star, and scaling to the M-dwarf surface density, we estimate
$N_9= 0.25$ (a more detailed description of this calculation will appear in Parravano, McKee
\& Hollenbach, in preparation).   

There are no direct observational estimates of the surface density of black holes, so we infer that
in an analogous manner to the computation of the number of neutron stars. 
We use the PMH11 IMF and the Hurley
et al (2000) evolution code, which relates the mass of the progenitor to the likelihood
and mass of the resultant black hole, and we scale the surface density of black holes
 to the local M dwarf surface density.
 In order to estimate the volume density of black holes in the midplane, we assume that the scale height
 is similar to that of M dwarfs; the actual value of the scale height is unknown.

\subsection{Results and Comparison with Previous Work}
\label{sec:comp}

The results on the local stellar surface densities are collected in Table 1.
We have estimated errors for the M-star surface density (13\% from Z01) and for the
surface density of white dwarfs (17\%, which is dominated by the uncertainty in $\SSM$).
We lump all the other visible stars together and assign an error of 15\%; for the 
brown dwarfs, neutron stars, and black holes, we assign an
error of 30\%. The resulting error in the total surface density is $\simeq 3\,M_\odot$~pc\ee\ and is dominated by the uncertainty in the surface density of main sequence stars.
Our final value for the local total surface density of stars and stellar remnants by mass is then
\beq
\Sigma_*=33.4\pm 3\ M_\odot~\mbox{pc\ee},
\eeq
whereas the local surface density of visible stars (main-sequence stars and giants) is
\beq
\Sigma_\vis=27.0 \pm 2.7 \,M_\odot~\mbox{pc\ee}.
\eeq
The surface density of stars and stellar remnants within 1.1 kpc of the Galactic plane is
\beq
\SSopo=31.2\pm 3 \,M_\odot~\mbox{pc\ee},
\eeq
where we have approximated the error as being the same as that for $\Sigma_*$.

These results are very close to those of F06: Our results for $\Sigma_*$ and $\SSopo$ 
are 8\% and 3\% less than theirs, respectively, whereas our result for $\Sigma_\vis$ is
1\% more than theirs.
However, the excellent agreement between our results and those of F06 is fortuitous, 
since the mass is distributed somewhat differently,
but in such a way that the sum of all the components is about the same.  We find there to be less mass in
main-sequence stars earlier than M, in
brown dwarfs, and in white dwarfs, but more mass in M dwarfs.
We have discussed
the reasons for these differences in Sections \ref{sec:revise} (M dwarfs), \ref{sec:msg} (more massive stars), \ref{sec:bd} (brown dwarfs), and \ref{sec:wd} (white
dwarfs).  This different distribution in the mass, especially that of M dwarfs and white dwarfs, is one of the main results of this paper.

Our result for the local surface density of visible stars is also within the errors of the results
of Bovy et al. (2012b), who found $\Sigma_\vis=30\pm 1\;M_\odot$~pc\ee\ for a Chabrier (2001) log-normal IMF.
Their result has some dependence on the IMF; for example, they
find $\Sigma_\vis=29\,M_\odot$~pc\ee\ for a Chabrier (2003) IMF. They did not take binarity into account in their analysis, but
Bovy \& Rix (2013) show that binarity would not affect their estimates of the surface density of visible stars within 1.1 kpc. Furthermore, we note that their result is based on observations of G stars,
which cover a narrow range of mass. As a result, relatively small errors in the mass-luminosity relation 
can alter the fraction of the stellar mass that is in G stars and thus make a significant contribution to the error
in the total mass of stars. Their quoted error does not allow for this effect (Bovy, private communication).

Finally, we note that the midplane density of stars is
\beq
\rho_{*0}=0.043\pm 0.04~~M_\odot\mbox{ pc\eee},
\eeq
where we have estimated the error to be 10\%; this is considerably larger than the 4\% difference between our result and that of F06, but there is a substantial overlap in the data.
The effective scale height of the stars is then
\beq
h_*=\frac{\Sigma_*}{2\rho_{*0}}=388\pm 40\mbox{  pc},
\label{eq:hstar}
\eeq
where we have again adopted an error of 10\%, which is the difference between the scale height found by \juric\ (2008) et al. and that found by Z01.

\section{The Local Interstellar Gas}
\label{sec:gas}

Having determined the densities of stars and stellar remnants in the solar vicinity, we turn now to a discussion
of gas in the local interstellar medium--i.e., within a cylindrical radius of 1 kpc of the Sun. 
We consider the molecular (\htwo), atomic (HI) and ionized (HII) gas in turn.
Read (2014) and Hessman (2015) have emphasized the uncertainties in existing determinations of the amount of
local interstellar gas, and our goal in the present section is to reduce those uncertainties.

\subsection{Molecular Gas: \htwo}

In their review, Heyer \& Dame (2015) state that the best value for the local surface density of \htwo\ remains that of
Dame et al. (1987), who found that the distribution 
of molecular gas within 1 kpc could be described by a mean midplane density
of H nuclei
$\nbh({\rm H_2})=0.2$~cm\eee\ 
and a half-width half-maximum of 87 pc. Their density estimate was based on a CO to \htwo\
conversion factor of $2.7\times 10^{20}$ cm$^{-2}$/(K km s$^{-1})$; using
the value of $2\pm 0.6\times 10^{20}$ cm$^{-2}$/(K km s$^{-1}$) suggested by Bolatto, Wolfire, \& Leroy (2013)
reduces the central density to 0.15 cm\eee. (It should be
noted that this empirical conversion factor includes the H$_2$ that lies in surface layers that have little CO
but mostly C$^+$, the so-called ``dark" or ``hidden" H$_2$.)
Dame et al. (1987) pointed out that it was not possible to determine
the shape of the spatial distribution of the local molecular gas because the number of massive molecular clouds,
which contain most of the mass, is too small. They adopted a Gaussian distribution 
$\propto \exp-(z/105~\rpc)^2$ based on observations
of more distant molecular gas by others. The resulting properties for the local molecular gas are given in Table
\ref{tab:ism}. In converting from the column density of hydrogen to the surface density of mass, we have assumed
that the mass in helium and heavier elements is 40\% of that in hydrogen, so
that the mass per hydrogen atom is $1.4\times m_\H=2.34\times 10^{-24}$~g (we shall henceforth refer to this as the
He correction).
 All of this is close to the midplane, at $z<1.1$ kpc.
 The uncertainty is dominated by the CO-to-\htwo\ conversion, which is a factor 1.3.

\subsection{Atomic Gas: HI}

Based on the Hat Creek Survey of the 21 cm emission in the Galaxy (Heiles \& Habing 1974), Heiles (1976) determined the
azimuthally-averaged column density
of neutral atomic hydrogen at Galactic latitude $b$ for $|b|\geq 10^\circ$ and for declinations $\delta>-30^\circ$
over the velocity range $-92$~km~s\e $<v<75$~km~s\e. 
Summing his results for gas above and below
the Galactic plane, we find $N_\HI^{\rm thin}(|b|)= [ (7.45\pm0.06)\csc |b| -(2.39\pm0.14) ]\times 10^{20}$~cm\ee,
where the superscript ``thin" indicates that the gas is assumed to be optically thin.
These observations were contaminated by stray radiation--i.e., radiation from directions other than that in which the telescope was pointed.
Heiles et al. (1981) addressed this problem by reobserving some areas of the sky with the horn reflector antenna at Bell Telephone Laboratories, which
has much less stray radiation than the Hat Creek telescope. Based on their results, we infer
that the column density of optically thin HI summed over latitudes $+b$ and $-b$ is
\beq
N_\HI^{\rm thin}(|b|)= [ (7.45\pm0.06)\csc |b| -(3.72\pm0.14) ]\times 10^{20}~~~\mbox{cm\ee}.
\label{eq:nhpt1}
\eeq
We shall refer to this as the ``Heiles model" for short.
Heiles et al. (1981) interpreted the second term as being due to the fact that the Sun is in a low-density region created
by stellar winds and supernovae. Indeed, the Sun is located in the Local Bubble, which has a density that is much
less than the average interstellar value (Cox \& Reynolds 1987).
We can then write
\beq
N_\HI^{\rm thin}(|b|)=\nhpt\csc |b| - N^{\rm thin}_\LB,
\label{eq:nhpt2}
\eeq
where $\nhpt$ is the column density of optically thin HI through the entire plane of the Galaxy 
outside the Local Bubble and $N^{\rm thin}_\LB$ is the average 
deficit of optically thin HI across the diameter of the Local Bubble. In reality, the Local Bubble is very aspherical,
so after averaging over Galactic longitude, $N^{\rm thin}_\LB$ should depend on Galactic latitude; this dependence is
ignored in the Heiles model. The average radius of the Local Bubble is $R_\LB=N^{\rm thin}_\LB/(2n_0^{\rm thin})=60/n_0^{\rm thin}$~pc, 
where $n_0^{\rm thin}$ is the midplane density of optically thin HI. 
The total mass surface density of gas, including He, corresponding to $\nhpt=7.45\times 10^{20}$~cm\ee\ is
$\Sigma^\thin=8.34\, M_\odot$ pc\ee.

We can check the accuracy of the Heiles model for the HI in the solar neighborhood based on the results of
Lockman \& Gehman (1991; hereafer LG91) from the Bell Laboratories HI Survey (Stark et al. 1992). LG91
provide values of $N_\HI^{\rm thin}$ weighted by sin$|b|$ for $|b|\geq (45^\circ,\,80^\circ)$ and $\delta\geq -40^\circ$. 
Our comparison is approximate: the Bell Laboratories HI Survey covers a broader range of velocities than 
the Hat Creek Survey, and in our comparison we assume that the surveys cover most of the sky with $|b|\geq 45^\circ$.
Weighting the Heiles model by $\sin|b|$ gives a prediction for what Lockman \& Gehman (1991) found:
\beqa
\avg{N_\HI^{\rm thin}(|b|)\sin|b|}&=& \frac{1}{\Delta\Omega}\int_{\Delta\Omega} N_\HI^{\rm thin}\sin(|b|) d\Omega,\\
&=&\nhpt-\frac 12 (1+\sin |b|)N_\LB^{\rm thin}.
\eeqa
For $|b|\geq 80^\circ$, LG91 find $\avg{N_\HI^{\rm thin}(|b|)\sin|b|}=3.2\times 10^{20}$~cm\ee\ (note that the values they cite are for
half the disk); the Heiles model gives $3.75\times 10^{20}$~cm\ee, 17\% larger. For $|b|\geq 45^\circ$, LG91 find
$\avg{N_\HI^{\rm thin}(|b|)\sin|b|}=4.2\times 10^{20}$~cm\ee, in excellent agreement with the prediction of the Heiles model, $4.3\times 10^{20}$~cm\ee.
These results suggest that the Heiles model for optically thin HI is accurate to within about 20\%.

In a separate study, Lockman (1984) studied the HI in the inner Galaxy and concluded that three components were needed
to describe the vertical structure, two Gaussians and one exponential; the exponential was required to
account for gas high above the plane (see below). 
Kulkarni \& Heiles (1987) identified the component with the smallest scale height
with the Cold Neutral Medium (CNM) and the other two components with the Warm Neutral Medium (WNM;
we label these WNM$_1$ and WNM$_2$). 
In their review, Dickey and Lockman (1990) updated the properties of the three components. The total surface density in the Dickey-Lockman
model, $\nhpt=6.2\times 10^{20}$ cm\ee, is less than that directly measured outside the local ``hole" by Heiles (1976),
$\nhpt=7.45\times 10^{20}$~cm\ee. It is no surprise that these numbers are different: The Dickey-Lockman
result applies to Galactocentric radii of 4--8 kpc, whereas the Heiles result is for the solar vicinity. Nonetheless, the
two numbers agree to within 20\%.
More recently, Kalberla \& Dedes (2008) found
a mass surface density at the solar circle
of $\Sigma^{\rm thin}=10\,M_\odot$~pc\ee.  Kalberla (private communication) has stated that this includes He, and that, furthermore,
a more refined estimate gives $\Sigma^{\rm thin}=9\,M_\odot$~pc\ee. This
corresponds to $\nhpt=8.0\times 10^{20}$~cm\ee, which is within 10\% of the Heiles model. Since the Heiles
model is intermediate between the Dickey-Lockman  results and Kalberla's, we shall adopt it as our fiducial
model.

In order to infer the vertical distribution of the optically thin HI as well as the total column, we 
adopt the three components found by Dickey \& Lockman (1990), but multiply the densities 
by 7.45/6.2 to bring the total column density up to the Heiles value.
The central densities of the CNM and WNM in this model are 0.475 cm\eee\ and 0.206 cm\eee;
the effective scale heights, $h_\eff\equiv N_\HI^{\rm thin}/(2n_0)$ are 113 pc and 338 pc, respectively.
These values for the central densities are quite close
to those found by Kalberla \& Dedes (2008), 0.50 cm\eee\ and 0.19 cm\eee, but the scale heights
are somewhat smaller than they found, 150 pc and 400 pc.

The total surface density of local HI in this model, $\Sigma_\HI^{\rm thin}=6.0\,M_\odot$~pc\ee, is somewhat less than the maximum
of about $10\,M_\odot$ observed in star-forming disk galaxies (Wong \& Blitz 2002). This is reasonable, since
the maximum is reached in regions that have significant amounts of molecular gas, whereas
the local ratio of \htwo\ to optically thin HI is only about 0.12. The surface density including the He and heavier elements
is $8.3\,M_\odot$~pc\ee. This is essentially the same as the $8\,M_\odot$~pc\ee\ given by Kulkarni \& Heiles (1987) after
allowing for the difference in the He correction (they used 1.36 instead of our 1.40); this is to be expected, since
both values are based on Heiles (1976). We estimate the uncertainty in the surface density of
the WNM as $\sim10$\% based on the magnitude of the stray radiation correction found by Heiles et al. (1981)
at a typical point in the sky ($|b|\sim 30^\circ$).
The CNM could have a larger error, since less than half the CNM within 1 kpc is in the volume with
$|b|>10^\circ$ with the density distribution in Table \ref{tab:ism}; we adopt 20\%, including optical depth
effects (see below). Since the columns of CNM and WNM are comparable
and the errors could well be correlated,
the total uncertainty in the HI column is about 15\%

\subsubsection{Correction for Optical Depth}

The amount of HI given above is a lower limit since it does not allow for the opacity of the CNM,
which causes the actual value of the HI column, $N_\HI$, to exceed the value inferred based on
the assumption that the HI is optically thin, $N_\HI^{\rm thin}$.
Strasser \& Taylor (2004)
found that $\calr\equiv N_\HI/N_\HI^{\rm thin}=1.32$ based on the Canadian Galactic Plane Survey,
which covered Galactic longitudes from $74^\circ$ to $147^\circ$ and Galactic latitudes from $-3.6^\circ$ to
$5.6^\circ$. 
Braun (2012) has studied the HI in M31, M33 and the LMC and concluded that $\calr=1.34\pm 0.05$ averaged over the entire galaxy. This value is in excellent agreement with that of Strasser \& Taylor (2004), although it must be borne in mind that the latter applies to gas close to the plane of the Galaxy, whereas Braun's result applies to the entire galaxy. Braun (2012) attributed
the self-absorption to large atomic clouds of order 100 pc in size.

We can estimate the value of $\calr$ for the solar vicinity from the results of Heiles \& Troland (2003).  
Their data show that there is little absorption for $|b|>40^\circ$
(in fact, there is very little CNM for $b>40^\circ$),
so we focus on
the latitude range $|b|<40^\circ$. From Fig. 5 in their paper, we find that
the 43 sources in this range have a mean value $\avg{\calr}=1.29$. They note that the line profiles for $|b|<10^\circ$
are complicated, which contributes to the uncertainty in this value. Unfortunately, there is no way to estimate the
magnitude of this uncertainty; what we can conclude is that there is no evidence that the local value of $\calr$ differs
significantly from the global values determined by Strasser \& Taylor (2004) and by Braun (2012).
All these values are consistent with $\calr=1.30$, and we adopt that value here. 
Note that $\calr$ includes HI that is associated with partially molecular gas.

Since it is only the CNM that is optically thick, we can determine the value of $\calr$ in the CNM, 
$\calr_\cnm=N_\cnm/N^\thin_\cnm$, from
\beq
\calr=\frac{\calr_\cnm N_\cnm^{\rm thin}+N_\wnm}{N_\HI^{\rm thin}}.
\label{eq:calr}
\eeq
In terms of the column density fraction $f_\wnm=N_\wnm/N^\thin_\HI$, this implies
\beq
\calr_\cnm=\frac{\calr-f_\wnm}{1-f_\wnm}.
\eeq
The Dickey-Lockman (1990) model has $f_\wnm=0.56$; for $\calr=1.30$, this implies that the CNM column densities are
larger than the observed values by a factor $\calr_\cnm=1.68$.

 Our results for the local distribution of HI are summarized in Table \ref{tab:ism}.
 We have used
 $\calr_\cnm=1.68$ to correct the observed values of the density and surface density of the CNM for optical depth
 effects.
Since the three estimates we have for $\calr$ are in good agreement, the uncertainty in $\calr$ should
not add significantly to the 20\% error we have already estimated for the
CNM.

Recently, Fukui et al. (2015) have analyzed the all-sky LAB 21 cm survey (Kalberla et al. 2005) together with
the {\it Planck} map of of the dust emission at 353 GHz ({\it Planck} Collaboration 2014) 
to infer the total amount of HI at $|b|>15^\circ$. A strength of their analysis is that the dust emission traces {\it all}
the gas, including opaque HI. They reach two controversial conclusions: First, they infer that the
``dark gas," which gamma-ray observations show is not associated with either CO emission or 21 cm emission
(Grenier et al 2005),
consists primarily of opaque HI; by contrast, other
observations (Lee et al. 2015 and references therein) 
and theory (Wolfire et al. 2010) are consistent with 
this gas consisting primarily of \htwo\ and atomic C. We note that the data they cite on direct measurements of
\htwo\ absorption show that there is indeed little \htwo\ in regions in which $N_\HI\la 1.0\times 10^{21}$~cm\ee; however,
for the two cases in which the column density is larger, the \htwo\ fraction is significant.
Second, and of direct relevance to the present study,
they conclude that typically $\calr\simeq 2-2.5$, suggesting that a significant amount of HI has been missed.
However, this contradicts studies of HI absorption: Whereas Fukui et al. (2015) find a median value of the HI optical
depth of order unity, Heiles \& Troland (2003) find a median value less than 0.2 (Fig. 5 in their paper shows that
the median value of $R_{\rm raw}=1/\calr$ exceeds 0.9, corresponding to $\tau_\HI\leq 0.21$). In their study
of the Perseus molecular cloud, Stanimirovi\'c et al. (2014) find a median value of the maximum optical depth
of $\tau_\HI=0.16.$ The reason for this discrepancy is not clear.
Reach et al. (2015) have confirmed the presence of
significant amounts of dust that does not appear to be associated with CO or with HI emission in a set of interstellar clouds with
masses of order $100\,M_\odot$. In order to account for the large amounts of dark gas in these clouds by optically thick HI, optical depths $>3$ are required, which are rare. On the other hand, they show that a model in which the dust is associated with \htwo\ agrees with the the observed dust temperatures. They also discuss a third possibility, that the observations are due to unusual dust
rather than the presence of dark gas, and conclude that is unlikely.

Regardless of the interpretation of Fukui et al.'s (2015) results, we find that they are in very good agreement with the Heiles model.
Extrapolation of their results for the relation between dust emission and 21 cm emission for the
warmest dust to zero 21-cm emission gives a negligible dust optical depth; we conclude that the ionized gas (Section \ref{sec:hii}) makes
a negligible contribution to the dust. They excluded regions with obvious CO emission from their analysis, so the gas
they analyze should consist of HI and dark \htwo\ that is not along the line of sight to CO emitting clouds.
Locally, all the \htwo\ is only a small fraction of the HI, and if the
\htwo\ along the line of sight to observable CO is excluded, 
the fraction is even smaller. We therefore expect gas observed by Fukui et al (2015) to be primarily
HI. Under the assumption that the gas is located at a distance
of 150 pc, their
results correspond to a total mass of gas 
(including He)
at $|b|\geq 15^\circ$ of $1.4\times 10^6\,M_\odot$ (Fukui 2015, personal communication). 
This corresponds to an average column density of HI as observed
along a line of sight of $\avg{N_{\HI,\,\los}}(|b|\geq 15^\circ)=6.0\times 10^{20}$~cm\ee\ (note that whereas $N_\HI$ is the column
density through the entire disk, $N_{\HI,\,\los}$ is measured from the Sun and goes through only half the disk).
The average column density 
of optically thin HI in the Heiles model is $\avg{N_{\HI,\,\los}}^\thin(|b|\geq 15^\circ)=4.9\times 10^{20}$~cm\ee.
Since the WNM has a larger scale height than the CNM, gas with $|b|\geq 15^\circ$ has a lower fraction of CNM
than the ISM as whole; hence, using $\calr=1.3$ to correct for optical depth gives an upper limit on the average
column density of HI,
$\avg{N_{\HI,\,\los}}(|b|\geq 15^\circ)<6.4\times10^{20}$~cm\ee. The average HI column density at
$|b|\geq 15^\circ$ inferred from the {\it Planck} data by Fukui et al. (2015) thus agrees with the range of values in the
Heiles model to within about 10\%. This good agreement implies that our estimate for the mass of neutral gas 
(mostly HI in the interpretation of Fukui et al. 2015,
but possibly including a significant amount of \htwo\ in other interpretations, such as that of Reach
et al. 2015) in this part of the sky (excluding the small fraction with detectable CO emission)
is reasonably accurate.
However, the agreement does not explain why the Heiles model is dominated
by optically thin HI whereas the analysis of Fukui et al. (2015) implies that a significant fraction
of the HI is opaque.

\subsubsection{HI in the Galactic Halo}

For purposes of discussion, we define the gaseous Galactic halo 
in the solar vicinity to start 1 kpc from the midplane. The Dickey-Lockman model gives $N_\HI(>1\;$kpc)$\,=1.3\times10^{19}$~cm\ee; our renormalized version of this model gives $1.6\times 10^{19}$~cm\ee. 
Lockman, Hobbs \& Shull (1986) provided some support for 
the existence of HI far from the plane by comparing Lyman$-\alpha$ absorption observations of stars far from
the plane with 21 cm emission observations in the same direction. They found that the HI column more than 1 kpc from
the plane in the solar vicinity is $(5\pm3)\times 10^{19}$~cm\ee. It should be noted that this result is less than 2 sigma,
and that observations of 5 of the 6 stars they observed more than 1 kpc from the plane were consistent with no
HI in the halo. Their result is thus consistent with the lower values in the Dickey-Lockman
and revised Dickey-Lockman models.

An independent estimate of the amount of HI in the local halo comes from 21-cm observations of high-velocity gas near
the Galactic poles. Kulkarni \& Fich (1985) found that up to 20\% of the HI near the poles could be in a component
with a velocity dispersion of 35~km~s\e. Lockman \& Gehman (1991) modeled observations of the polar regions
with hydrostatic equilibrium models and found that they needed a high velocity-dispersion component with
$\sigma=28,\,34$~km~s\e\ and
an implied effective scale height $h_\eff=N_\HI/(2n_0)=630,\,770$ pc for the two models they considered.
If we approximate the vertical distribution of gas with an exponential of this scale height, then their results
imply $N_\HI(>1\;$kpc)$\,\simeq 1.4\times 10^{19}$~cm\ee, where we have included the small contribution from
the other components in their model. This result is thus consistent with both the original Dickey-Lockman model
and our revision of it.

Kalberla et al. (1998) found a larger amount of gas, with a total column density of $2.8\times 10^{19}$~cm\ee, 
with a higher velocity dispersion, $\sigma=60$~km~s\e. They estimated the scale height of this gas as 4.4 kpc,
so the amount more than 1 kpc from the plane is $2.2\times 10^{19}$~cm\ee. This is somewhat larger than our
estimate of $1.6\times 10^{19}$~cm\ee, but the additional gas is more than 6 kpc from the plane.

Our renormalized Dickey-Lockman model gives a column density of $1.25\times 10^{19}$~cm\ee\ above 1.1 kpc,
corresponding to a mass surface density of $0.14\,M_\odot$~pc\ee. This is a small correction, so the surface
density of HI within 1.1 kpc of the plane is very nearly equal to the total value (Table \ref{tab:ism}).

\subsection{Ionized Gas: HII}
\label{sec:hii}

The total HII surface density at the solar circle can be separated into 
photoionized gas at $\sim 10^4$ K and collisionally ionized gas at $\sim 10^6$ K (HIM). The photoionized gas can in turn
be divided into diffuse ionized gas (DIG or WIM) and HII regions and their envelopes.
Pulsar dispersion measures (DMs) provide a direct measurement of the column density of all the ionized gas. Since the HIM 
makes only a small contribution to the DM for $z\la 5$~kpc (Gaensler et al. 2008), we shall not treat it here.
The classical model for the distribution of electrons in the Galaxy is that by Taylor \& Cordes (1993), which
was refined by Cordes \& Lazio (2003). The dominant features in the solar vicinity in these models are a thick
disk with a scale height $\sim 0.9$~kpc and the Gum Nebula. Subsequently, Gaensler et al. (2008) used pulsars
with known distances to show that the scale height of the ionized gas is considerably larger, about 1.8~kpc.
Recently, Schnitzeler (2012) has reviewed these and other models for the electron distribution and
concluded that the model that best fits the observations is a revised Taylor-Cordes model  in which
the thick disk has a scale height
of 1.59 kpc and a DM normal to the plane of 24.4 cm\eee\ pc. Schnitzeler (2012) subtracted out the effects of
all other features in the Taylor-Cordes model other than the thick disk before determining the parameters
of the disk. The dominant such feature in the solar vicinity is the Gum Nebula, and based on the work
of Purcell et al. (2015), we estimate that this contributes about $0.11\,M_\odot$~pc\ee\ to the surface density
within a kpc of the Sun. We estimate that the mass of ionized gas in other features
in the solar vicinity, such as the Eridanus Bubble, have much smaller masses of ionized
gas. We present these results in Table \ref{tab:ism}, where we have
added the $0.11\,M_\sun$~pc\ee\ from the Gum Nebula to $N_\HI$, $\Sigma$ and $\Sigma_\opo$.
An estimate of the error comes from comparing Schnitzeler's DM with Gaensler et al.'s: they differ by only 5\%.
We have allowed for a somewhat larger fractional error in $\Sigma_\opo$ since the scale height is also uncertain.

\begin{deluxetable}{lcccc}
\tablecolumns{5}
\tablewidth{0pc}
\tablecaption{The Vertical Distribution of the Local Interstellar Medium\tnm{a}
\label{tab:ism}}
\tablehead{
\colhead{} & 
\colhead{$\bar n_\H(z)$}&
\colhead{$\nhp$}&
\colhead{~~$\Sigma_g$\tnm{b}~~}&
\colhead{~~$\Sigma_{g,1.1}$\tnm{b}~~} \\
\colhead{} & 
\colhead{(cm\eee)}&
\colhead{($10^{20}\,$cm\ee)}&
\colhead{~~($M_\odot$ pc\ee)~~} &
\colhead{~~($M_\odot$ pc\ee)~~} \\
\vspace{-0.4cm}
 }
 \startdata
  ~~H$_2$ &$0.15\exp-(z/105~\rpc)^2$&$0.9\pm 0.3$& $1.0\pm0.3$ & $1.0\pm0.3$   \\
   \vspace{-0.4cm}\\ 
  \hline
  \vspace{-0.4cm}\\
 ~~HI: CNM\tnm{c} &$0.80\exp-(z/127~\rpc)^2$&5.54&6.21&6.21  \\
 ~~~~~~\,WNM$_1$     & $0.13\exp-(z/318~\rpc)^2$ & 2.24 &2.51&2.51  \\
 ~~~~~~\,WNM$_2$   & $0.077\exp-(z/403~\rpc)$& 1.91& 2.14&2.00  \\
    \vspace{-0.4cm}\\
  \hline
  \vspace{-0.4cm}\\
 ~~~~~~\,Total HI &$n_{\rm HI,0}=1.01$ &$9.7\pm 1.5$&$10.9\pm1.6$&$10.7\pm1.6$\\ 
   \vspace{-0.4cm}\\
  \hline
  \vspace{-0.4cm}\\
 ~~HII\tnm{d}   & $0.0154\exp-(z/1590~\rpc)$&$1.6\pm0.1$&$1.8\pm0.1$&$ 0.9\pm0.1$ \\
 \vspace{-0.4cm}\\
 \hline
 Total &---&  $12.2\pm1.5$ &$13.7\pm1.6 $& $12.6\pm1.6$ 
\enddata
\tnt{a}{References:  \htwo: Dame et al. (1987), Heyer \& Dame (2015). HI: based on (see text) 
Heiles (1976), Dickey \& Lockman
(1990), Heiles \& Troland (2003). HII: Schnitzeler (2012).}
\tablenotetext{b}{Includes He and heavier elements with 40\% of the mass of H.}
\tnt{c}{Optically thin results for the CNM are $n_{\cnm,0}^{\rm thin}=0.48$~cm\eee, $
N^\thin_\cnm
=3.30\times 10^{20}$~cm\ee, $\Sigma^{\rm thin}_\cnm=\Sigma_{\cnm,\opo}^{\rm thin}=3.70\mpc.}
\tnt{d}{The values of $\nhp$, $\Sigma$, and $\Sigma_\opo$ include $0.11\,M_\odot$~pc\ee\ for the Gum Nebula that is not included in $\bar n_\H(z)$.}
\end{deluxetable}

\subsection{Local Interstellar Gas: Summary}

Table \ref{tab:ism} summarizes our results for the three different components of the gas.   Our results can be compared with the
results of Bahcall et al. (1992), which were adopted with only minor changes by F06. 
Bahcall et al. (1992) estimated $3\,M_\odot$~pc\ee\ for the molecular gas  based on the results of Scoville
\& Sanders (1987). These authors adopted a CO-to-\htwo\ conversion factor of $3.6\times 10^{20}$~cm\ee~(K~km~s\e)\e;
reducing that to the value recommended by Bolatto et al. (2013) 
reduces the column density of molecular gas to $1.7\,M_\odot$~pc\ee.
This is actually the value at the solar circle, not the solar vicinity, and since the Sun is in an interarm region it is
not surprising that the local value we estimated, $1.0\,M_\odot$~pc\ee, is smaller. The HI data used by Bahcall et al. (1992)
is based on Heiles (1976), just as ours is; we obtained $10.9\,M_\odot$~pc\ee\ instead of $8\,M_\odot$~pc\ee\ as they
did because we corrected for optical depth effects and because we used a slightly larger He correction.
We showed that a model for the distribution of local HI based on the work of Heiles and collaborators (the ``Heiles model")
produces a total amount of local gas at $|b|\geq 15^\circ$ that is in good agreement
with the results of Fukui et al. (2015) based on {\it Planck} data, implying that we are not missing a significant amount of
unseen dark gas.
For the ionized gas, Bahcall et al. (1992) adopted $2\,M_\odot$~pc\ee\ from Kulkarni \& Heiles (1987), very close
to the $1.8\,M_\odot$ we obtained from more modern data. Since our somewhat larger value for the HI column
compensates for our lower value of the \htwo\ column, we find a total surface density that is in good
agreement with theirs: $13.7\,M_\odot$~pc\ee\ vs. $13\,M_\odot$~pc\ee. However, this gas is often assumed to
be entirely within 1.1 kpc of the plane, whereas more modern data on pulsar dispersion measures shows that almost
half the ionized gas is more distant than that; as a result, we estimate that
the surface density of gas within 1.1 kpc of the plane is $12.6\,M_\odot$~pc\ee.
Finally, we note that Table \ref{tab:ism} implies that
the total midplane density of the local interstellar gas is
$\bar n_{\H0}= 1.17$~cm\eee, corresponding to $\rho_{g0}=0.041\mpce,
where we have summed the densities of each HI component rather than using the rounded value of 1.01~cm\eee\ to
obtain this result. This is quite close to the
value $\bar n_{\H0}=1.2$~cm\eee\ obtained by Spitzer (1978), which was obtained by direct observations of gas in the plane.
The uncertainty in this estimate is about 10\%.
The effective scale height of the gas is
$h_g=\Sigma_g/(2\rho_{g0})\simeq 170$~pc.

\section{The Local Baryon Budget}
\label{sec:budget}

\begin{deluxetable}{lcccccc}
\tablecolumns{7}
\tablewidth{0pc}
\tablecaption{The Local Baryon Budget ($M_\odot$~pc\ee), Excluding Halo Stars at $z>3$ kpc
 \label{tab:baryon}}
\tablehead{
\colhead{} & 
\colhead{$\Sigma_*$}& 
\colhead{$\Sigma_{*1.1}$} &
\colhead{$\Sigma_g$}&
\colhead{$\Sigma_{g1.1}$}&
\colhead{$\Sigma_b$\tnm{a}}&
\colhead{$\Sigma_{b1.1}$\tnm{a}}
}
\startdata
\multicolumn{7}{c}{Star Counts}\\
\vspace{-0.4cm}\\
\hline
\vspace{-0.4cm}\\
Flynn et al. (2006)\tnm{b}&35.9\tnm{c}&32.0\tnm{d}&13.2&12.5\tnm{d}&49.1&44.5\tnm{d}\\
Read (2014)&$37.2\pm 1.2$ &
$34.7\pm 1.1$\tnm{e}&$17.0\pm 4.4$&$16.3\pm 4.2$\tnm{d}&$54.2\pm 4.9$&
$51.0\pm 4.6$\tnm{e}\\
This work&$33.4\pm3$&$31.2\pm3$&$13.7\pm 1.6$&$12.6\pm 1.6$&$47.1\pm 3.4$&$43.8\pm 3.4$\\
\cutinhead{Stellar Dynamics}
Kuijken \& Gilmore (1989b)&$35\pm 5$&---&$13\pm 3$&---&$48\pm 8$&---\\
Garbari et al. (2012)&---&$33.4^{+5.5}_{-5.2}$&---&$12_{-2.0}^{+1.9}$&---&$45.5_{-5.9}^{+5.6}$\\
Bovy \& Rix (2013)&$38\pm 4$\tnm{f}&---&13\tnm{g}&---&$51\pm 4$&---\\
Zhang et al. (2013)& ---&$42\pm 5$\tnm{d}&---&13\tnm{g}&---&$55\pm 5$\\
Bienaym\'e et al. (2014)&---&---&---&---&$44.4\pm 4.1$&$43.3\pm4.0$\tnm{d}\\
 \hline
\enddata
\tnt{a}{Errors are taken from the original reference where possible, and do not always correspond to summing the errors in $\Sigma_*$ and $\Sigma_g$
in quadrature.} 
\tnt{b}{Mixed approach: Stellar dynamics used to infer $\Sigma_*$ for main sequence stars with $M_V<8$.}
\tnt{c}{Includes $0.4\mpc\ for halo stars within 3 kpc of the plane.}
\tnt{d}{Our inferred value based on data in reference.}
\tnt{e}{Estimated value at 1.1 kpc based on our result, $\Sigma_{*1.1}=0.934\Sigma_*$.}
\tnt{f}{The stellar surface density is the total value, not the value within 1.1 kpc as implied by the text (Bovy, private
communication).}
\tnt{g}{Taken from F06; Zhang et al. (2013) assumed that this applies to gas within 1.1 kpc of the plane.}
\end{deluxetable}

Our results for the stellar surface density, $\Sigma_*$, gas surface density, $\Sigma_g$, and
total baryonic mass surface density, $\Sigma_b$, in the solar vicinity are compared to those of
others in Table \ref{tab:baryon}. Total values (excluding stars at $z>3$~kpc) and values within
1.1 kpc of the Galactic plane are given. 
We have included the results of Kuijken \& Gilmore (1989b), F06, and
all the ``latest measurements" summarized by Read (2014), except for the following: Moni Bidin et al
(2012b), whose work was criticized by Bovy \& Tremaine (2012); Bovy \& Tremaine (2012), who investigated only
the dark matter; and Smith et al. (2012), who deliberately presented an over-simplified toy model in which
all the matter was divided into a thin sheet and a uniform distribution. 
We include the study of  Bienaym\`e et al. (2014), which appeared
after the Read (2014) review.
We have divided the references into two groups: The first group relies on
star counts to infer the surface densities
of the M dwarfs and stellar remnants (F06, this work) or for all the stars (Read 2014). In the latter work, Read (2014)
adopted $\Sigma_{*,\MS}=30\pm1\;M_\odot$~pc\ee\ from Bovy et al. (2012b) and took the surface
density of stellar remnants and brown dwarfs in the thin disk from F06; he did not 
include the stellar remnants and brown dwarfs in the thick disk. The
second group inferred the stellar surface density by using stellar dynamics to interpret observations of 
the stellar velocities of a tracer population.

Our results for $\Sigma_*$ and $\Sigma_b$, both total and within 1.1 kpc,
are consistent within 1 sigma with those of other workers with the exception of Read (2014) and Zhang et al.
(2013). 
In inferring $\Sigma_{b1.1}$ from Bienaym\'e et al. (2014), we adopted
their value for the scale height of 300 pc; had they used our results for the vertical distribution of baryons, the difference
between their values of $\Sigma_b$ and $\Sigma_{b1.1}$ would have been larger.
The discrepancy with Read (2014) is primarily due to the large gas surface density he adopted based on
his assumption that the Kalberla \& Dedes (2008) HI column density did not include He, whereas it did (see above).
The outlier among these works
is Zhang et al. (2013), who find $\Sigma_b=55\pm 5\;M_\odot$~pc\ee. They provide only $\Sigma_{*1.0}$, 
and we have made the conservative assumption
that this is the same as $\Sigma_{*1.1}$; if the vertical distribution of baryons follows our model instead, then
their value for $\Sigma_{*1.1}$ would be $56\, M_\odot$~pc\ee.

Summing the densities of the gas (Table \ref{tab:ism}) and the stars (Table \ref{tab:local}), we find that the total
midplane density of baryons implied by our results is
\beq
\rho_{b0}=0.084\pm0.012\;M_\odot\;\mbox{pc\eee},
\label{eq:rhobo}
\eeq
where we have assumed an error of 15\%, somewhat larger than that for the surface densities.
This value of the midplane baryon density
is somewhat less than (but within the errors of) the value of $0.091\pm 0.009\, M_\odot$~pc\eee\ found by F06, primarily because their
model of the interstellar gas gives a higher midplane density ($\bar n_{\H0}=1.45$~cm\eee\ vs our value of 1.17 cm\eee).
Our value 
of $\rho_{b0}$ is within the errors of
that of Bienaym\'e et al. (2014), who found $\rho_{b0}=0.077\pm 0.007\,M_\odot$~pc\eee, 
and of Garbari et al. (2012), who found $\rho_{b0}=0.098^{+0.006}_{-0.014}\, M_\odot$  pc\ee.

\section{The Local Density of Dark Matter, $\rho_\dm$}

\begin{deluxetable}{lccccc}
\tablecolumns{6}
\tablecaption{Local Density of Dark Matter
\label{tab:dm}
}
\tablehead{
\colhead{~~Reference~~}&
\colhead{~~$\Sigma_{z}$~~}&
\colhead{~~$\Sigma_{b,z}$~~}&
\colhead{~~$\rho_\dm$}&
\colhead{~\vline~~}&
\colhead{$\rho_\dm$~(This work)\tnm{a}}\\
\colhead{ }&
\colhead{$(M_\odot$ pc\ee)}&
\colhead{$(M_\odot$ pc\ee)}&
\colhead{$(M_\odot$ pc\eee)}&
\colhead{ ~\vline~~}&
\colhead{ $(M_\odot$ pc\eee)}
}
\startdata
Kuijken \& Gilmore (1991)&$\Sigma_{1.1}=73.5\pm 6$\tnm{b}&$\Sigma_{b1.1}=48\pm 8$& 0.010\tnm{c}&\vline & $0.0135\pm0.0031$\\
Bovy \& Rix (2013)&$\Sigma_{1.1}=68\pm 4$&$\Sigma_{b}=51\pm 4\tnm{d}$ & 0.008&\vline&  $0.0110\pm0.0024$\\
Zhang et al. (2013)&$\Sigma_{1.0}=69\pm 6$\tnm{b}&$\Sigma_{b1.0}=55\pm 5$&0.0065&\vline&$0.0130\pm0.0034$\\
Bienaym\'e et al. (2014)\tnm{e}&$\Sigma_{1.0}=70.5\pm1$\tnm{b}&$\Sigma_{b1.0}=43\pm4$ & 0.0143&\vline&$0.0137\pm 0.0017$\\
\enddata
\tablenotetext{a}{From Equation (\ref{eq:rhodm1}) using the total surface density from the reference and the baryonic surface densities found here,
$\Sigma_{b1.1}=43.8\pm 3.4\mpc\ and $\Sigma_{b1.0}=43.1\pm 3.3\mpc.}
\tnt{b}{Inferred from value of $K_z/(2\pi G)$ given in reference using Equation (\ref{eq:sigkz}) with $\alpha=0$.}
\tnt{c}{Inferred from their values for  $\Sigma_{1.1}$ and $\Sigma_{b1.1}$.}
\tnt{d}{The authors neglected the difference between $\Sigma_b$ and $\Sigma_{b1.1}$ (Bovy, private communication).}
\tnt{e}{For their scale height of 300 pc, $\Sigma_{b1.0}\simeq\Sigma_{b1.1}\simeq 43\mpc.}
\end{deluxetable}

It is now possible to combine our determination of the local surface density of baryons,
$\Sigma_{b}(z)$, with the measurements
of the total mass surface density, $\Sigma(z)$, by others to infer the local density of dark matter.
There is no definitive evidence for a thin disk of dark matter (Kuijken \& Gilmore 1989b), and existing data are
consistent with an approximately spherical dark matter halo (Read 2014). Here
we therefore assume that the density
of dark matter near the plane is constant, reserving discussion of a possible thin disk of dark matter to section \ref{sec:thin} below.
It follows that the density of dark matter near the plane is proportional to the difference between the surface density of all matter and
that of the baryons,
\beq
\rho_\dm=\frac{\Sigma(z)-\Sigma_b(z)}{2z}.
\label{eq:rhodm1}
\eeq

The determination of $\Sigma$ generally proceeds in two steps:
First, the Boltzmann equation or an equation derived from it, such as the vertical Jeans
equation, is used to analyze the kinematics of a tracer population of stars and thereby determine the vertical
gravitational acceleration due to all matter within a distance $z$ of the plane, $K_z$. Then Poisson's 
equation is used to relate $\Sigma$ to $K_z$.
As shown in Appendix \ref{app:kz}, this relation is
\beq
\Sigma=\frac{|K_z|}{2\pi G}+\Delta\Sigma,
\eeq
where
\beq
\Delta\Sigma\equiv\frac{1}{2\pi GR}\pbyp{R}\int_0^z v_c^2 dz,
\eeq
and where $v_c$ is the circular velocity at $(R,z)$. If the rotation curve is flat throughout the region of interest, then $\Delta\Sigma=0$ and 
$K_z$ is directly proportional to $\Sigma$; this approximation is often made. However, 
$v_c(R,z)$ generally varies with $R$ for $z\neq 0$ even if the rotation curve is flat at $z=0$ (Bovy \& Tremaine 2012).
As a result, $|K_z|\neq 2\pi G\Sigma$ even for a flat rotation curve.

In Appendix \ref{app:kz} we show that $\Delta\Sigma$ can be approximated as the sum of two terms. The first, $\Delta\Sigma_1$, depends on
the slope of the rotation curve:
\beq
\Delta\Sigma_1=\frac{\alpha z v_{c0}^2}{\pi GR^2},
\eeq
where 
\beq
\alpha\equiv\ppbyp{\ln v_{c0}}{\ln R}
\eeq
measures the slope of the rotation curve
and $v_{c0}$ is the rotation velocity in the midplane. This term vanishes for a flat rotation curve. Bovy et al. (2012a) find
$\alpha=0.01^{0.01}_{-0.1}$, which is consistent with a flat rotation curve. 
The second term, $\Delta\Sigma_2$, is independent of the shape of the rotation curve and is generally ignored (e.g., Kuijken \& Gilmore
1989a).
If the density in the disk (including the 
dark matter) is approximated as a double exponential, $\rho\propto\exp(-R/h_R)\exp(-z/h_z)$, then this term is
\beq
\Delta\Sigma_2=\frac{z^2}{2h_R^2}\left(1-\frac{h_R}{R}\right)\left[1-\frac{2h_z}{z}+\frac{2h_z^2}{z^2}\left(1-e^{-z/h_z}\right)\right]\Sigma_0.
\eeq
If the vertical distribution of matter were actually exponential, then $\Sigma_0$ would be the total surface density of matter; this
clearly does not work for the dark matter. We are interested in applying this equation at a distance $\sim 1$~kpc from the plane, which
includes most of the baryons and where the surface density is dominated by the baryons. We therefore approximate $\Sigma_0$ as
the total surface density within a distance $z\simeq (1-1.1)$ kpc of the plane, and in our numerical evaluations we set $\Sigma_0=70\mpc. 
As noted in Appendix \ref{app:kz}, this double-exponential approximation for the density and surface density is accurate to within
about 5\%.
Observe that $\Delta\Sigma_2$ is positive and is independent of
the rotation curve. We then have
\beq
\Sigma\simeq \frac{|K_z|}{2\pi G} +\left(\frac{v_{c0}^2z}{\pi GR^2}\right)\alpha+\Delta\Sigma_2,
\eeq
which clearly shows the dependence of $\Sigma$ on the slope of the rotation curve.
In Appendix \ref{app:kz} we show that this
result agrees well with the numerical results of Bovy \& Rix (2013).
Now, $R=8.3$~kpc (Chatzopoulos et al. 2015) and $h_R=2.5$~kpc (Bovy \& Rix 2013). 
We adopt the standard value for the rotational velocity at the solar circle, $v_{c0}=220$~km~s\e, which is consistent with the recent
result of Bovy et al. (2012a), $v_{c0}=218\pm 6$~km~s\e.
We borrow results from below to infer
that $h_z=\Sigma_0/(2\rho_0)\simeq(70\mpc)$/(2\times 0.093\mpce)=0.376~kpc. As a result, we obtain
\beq
\Sigma=\frac{|K_z|}{2\pi G}+\left\{\begin{array}{l}5.20\alpha_{0.1}+2.00~~~M_\odot\mbox{ pc\ee}~~~(z=1.0\mbox{ kpc}),\\
	5.72\alpha_{0.1}+2.54~~~	M_\odot\mbox{ pc\ee}~~~(z=1.1\mbox{ kpc}),
	\end{array}\right.
	\label{eq:sigkz}
\eeq
where $\alpha_{0.1}\equiv\alpha/0.1$. 
It should be borne in mind that both terms in $\Delta\Sigma$ increase with height above the plane:
$\Delta\Sigma_1\propto z$ and  $\Delta\Sigma_2\propto z^2$ (approximately).

\subsection{Local Values of $\Sigma$ and $\rho_\dm$}

Read (2014) has summarized the determinations of $\rho_\dm$ through 2013. Here we use our results for
the baryon surface density, $\Sigma_b$, and for the relation between $\Sigma$ and $K_z$ (Eq. (\ref{eq:sigkz})) to revise the values
of $\Sigma$ and $\rho_\dm$ for the solar neighborhood obtained by a subset of these references.
We include the classic work of Kuijken \& Gilmore (1991). Among the latest local measures of $\rho_\dm$
cited by Read (2014), we
include Bovy \& Rix (2013) and Zhang et al. (2013). As noted in Section \ref{sec:budget}, we exclude
Moni Bidin et al. (2012b), Bovy \& Tremaine (2012), and Smith et al. (2012). 
In addition, we have omitted Garbari et al. (2012), since Read (2014) has shown that the total surface density of matter
inferred by their method is very sensitive to $\Sigma_b$.\footnote{Garbari et al. (2012)  find $\Sigma_{b1.1}=45.5\,M_\odot$~pc\ee\ 
and $\rho_\dm=0.022_{-0.013}^{+0.015}\,M_\odot$~pc\eee, which implies a total column density of matter of
$\Sigma_{1.1}=94\pm 31\,M_\odot$. On the other hand, Read (2014) shows that if one increases the surface density
of baryons to $55\,M_\odot$~pc\ee,
then $\rho_\dm$ drops from $0.022\,M_\odot$~pc\eee\ to $0.008\,M_\odot$~pc\eee, which corresponds
to $\Sigma_{1.1}=74\,M_\odot$~pc\ee. In other words, increasing $\Sigma_b$ by $10\,M_\odot$~pc\ee\ decreases
$\Sigma_{1.1}$ by $20\,M_\odot$~pc\ee, and it reduces the implied $\rho_\dm$ by almost a factor 3. We emphasize that
this in no way implies that their method is incorrect, only that we cannot use Equation (\ref{eq:rhodm1}) to infer the
value of $\rho_\dm$ for different values of $\Sigma_{b}$ since $\Sigma$ and $\Sigma_b$ are strongly coupled.} 
Two new determinations of $\rho_\dm$ have appeared since
Read's (2014) review, one by Bienaym\'e et al (2014), which we include, and one by Piffl et al. (2014). We do not
include the Piffl et al. work in our analysis since their determination of $\rho_\dm$ is far more accurate (15\% for a
spherical halo) than an estimate based on Equation (\ref{eq:rhodm1}).
We note that in many cases the distribution of matter is divided into a disk component, assumed to be baryonic,
and a uniform component, assumed to be dark matter, and the masses of these components are determined
from observations of stellar kinematics. However, as we have seen in Section \ref{sec:hii}, the scale
height of the ionized gas is sufficiently large (1590 pc) that it could be confused with the dark matter. Our method
does not suffer from this potential problem since we determine the surface density of the baryons directly.

Table \ref{tab:dm} presents the data for the four studies we have selected to analyze to determine $\rho_\dm$.
For Bovy \& Rix (2013), the value of $\Sigma$ is taken directly from the reference; the other three references actually provide
values of $K_z/(2\pi G)$, and we have converted to values of $\Sigma$ by using Equation (\ref{eq:sigkz}) under the
assumption that the rotation curve is flat ($\alpha=0$). Inserting these values of $\Sigma$ into Equation (\ref{eq:rhodm1})
with our values of $\Sigma_b$, we obtain the values of $\rho_\dm$ given in the right-hand column.

Our results can be summarized as
\beq
\rho_\dm=0.013\pm 0.003~~M_\odot\mbox{ pc\eee}= 0.49\pm0.13~~\mbox{GeV cm\eee},
\label{eq:rhodm2}
\eeq
where the uncertainty is the average of the uncertainties in Table \ref{tab:dm}.
This uncertainty is based on the assumption that there is not a separate component of dark matter in a thin disk
(see Section \ref{sec:thin} below).
These uncertainties have been calculated from the quoted uncertainties in $\Sigma$ given in the references
together with the uncertainties in $\Sigma_{b}$ from our work.
One can show that the error estimate in Equation (\ref{eq:rhodm2}) remains valid even if allowance is made for the uncertainty
in the slope of the rotation curve found by Bovy et al. (2012a), $-0.09<\alpha<0.02$.
This result is consistent with that of Bienaym\'e et al. (2014),
$\rho_\dm=0.0143\mpce,
but is more than
a standard deviation above the estimates of Bovy \& Rix (2013), $\rho_\dm=0.008\mpce,  and of Zhang et al. (2013),
$\rho_\dm=0.0065\mpce.
The discrepancy with Bovy \& Rix (2013) can be reduced to less than 1 sigma by allowing for the fact that the column density of baryons within 1.1 kpc is less than the total; if the reduction is the same as we found, $\Sigma_{b1.1}/\Sigma_b=0.931$, then the Bovy \& Rix value of $\Sigma_{b1.1}$ would drop to $47.5\,M_\odot$~pc\ee, and $\rho_\dm$ would increase to $\rho_\dm=0.0093\pm 0.0026\,M_\odot$~pc\eee.
Comparing our result with recent references that are not included in Table \ref{tab:dm}, we note that
our result is in excellent agreement with that of Piffl et al. (2014), $\rho_\dm=0.0126\pm 15\%\mpce, and it is consistent with
the results of Garbari et al. (2012), $\rho_\dm=0.022_{-0.013}^{+0.015}\,M_\odot$~pc\eee, and of
Smith et al. (2012), $\rho_\dm=0.015\,M_\odot$~pc\eee.

In our view, the cleanest determination of the local density of dark matter is that of Bovy \& Tremaine (2012,
hereafter BT12), who analyzed the data of Moni Bidin et al. (2012a,b) on stars far enough above the plane (1--4 kpc)
that the variation of gravitational acceleration with height is almost
entirely due to dark matter. They found $\rho_\dm=0.008\pm0.003\,M_\odot$~pc\eee. However, they then pointed
out that there were two effects that would increase the density in the plane above this: First, since the dark matter
was measured at an average distance of 2.5 kpc above the plane, the radial distance to the Galactic Center
is greater than in the midplane; in the NFW potential
(Navarro et al. 1997) they adopted, this means that the density is 7\% higher in the midplane than where it was measured.
Second, in the gravitational field of the disk, the dark matter density varies as $\exp(-\Phi/\sigma^2)$, where
$\Phi\simeq 2\pi G \Sigma_b z$ is the gravitational potential and $\sigma\simeq 130$~km~s\e\ is the velocity
dispersion of the dark halo.\footnote{If the disk formed suddenly, then the velocity dispersion of the dark matter near
the disk would be larger than this and there would be no overdensity. 
However, star formation appears to have proceeded steadily over the life of the disk (e.g., Binney et al. (2000), 
Just \& Jahrei{\ss} 2010), so this effect is unlikely to be important.}
Altogether, this makes the local density of dark matter larger than that 2.5 kpc above
the plane by a factor of about 1.3, so that
\beq
\rho_\dm(\mbox{BT12})=0.010\pm 0.004~~M_\odot\mbox{ pc\eee}.
\eeq
This is consistent with our result and with all the results in Table \ref{tab:dm}.

Read (2014) lists five global measures of the dark matter density based on fitting the rotation curve of the Galaxy
under the assumption of a spherical dark matter halo. Our results are consistent with all of these, although only
marginally in the case of Weber \& de Boer (2010), who found $\rho_\dm=(0.005-0.010)\mpce. The two references that
adopted the weakest priors are fully consistent with our result: Salucci et al. (2010) found $\rho_\dm=0.011\pm0.005\mpce,
and Iocco et al. (2011) found $\rho_\dm=(0.005-0.015)\mpce.

Finally, our estimate of the local matter density, both baryons and dark matter, in the plane is
\beq
\rho_0= 0.097\pm 0.013\, M_\odot\mbox{ pc\eee}.
\label{eq:rhoo}
\eeq
This result is compared with others in Table \ref{tab:oort}. We include two classical pre-{\it Hipparcos} values, one by
Oort (1960) and one by Kuijken \& Gilmore (1989); both are higher than our result. The highly cited work of Holmberg
\& Flynn (2000) has a value in good agreement with ours. Among Read's (2014) ``latest measurements," only 
Garbari et al. (2012) give a result for $\rho_0$; it is slightly more than one sigma above ours.
Our result is in good agreement with the
value obtained by Bienaym\'e et al. (2014), $\rho_0=0.091\,M_\odot$~pc\eee.

\begin{deluxetable}{lc}
\tablecolumns{2}
\tablecaption{The Oort Limit
\label{tab:oort}
}
\tablehead{
\colhead{~~Reference~~}&
\colhead{ ~~$\rho_0\;(M_\odot$ pc\eee)}
}
\startdata
Oort (1960)\tnm{a}&$0.15 \pm 10\%$\\
Kuijken \& Gilmore (1989c)\tnm{a,b}&$0.11 - 0.29$\\
Holmberg \& Flynn (2000) &$0.102 \pm 0.010$\\
Garbari et al. (2012)  & $0.120^{+0.016}_{-0.019}$ \\
Bienaym\'e et al. (2014)&$0.091 \pm 0.006$\\
This work & $0.097 \pm 0.013$\\
\enddata
\tnt{a}{Pre-{\it Hipparcos} estimates.}
\tnt{b}{The intent of this work was to show that the estimation of the Oort limit depends on the star sample analyzed; 
the lower value corresponds to the estimate using a subsample of F8 dwarfs, whereas the upper one corresponds to a 
subsample of F5 dwarfs. }
\end{deluxetable}

\subsection{A Thin Disk of Dark Matter?}
\label{sec:thin}

Under the assumption that the density of dark matter near the plane is constant, we have found that the local ratio of
the total density of matter to that of the baryons is $\rho_0/\rho_{b0}=0.093/0.08=1.16$. Historically, some workers have
found significantly larger values: for example, Oort (1960) found $\rho_0/\rho_{b0}=2$ (although stars fainter than $M_V=15$ were
not included in $\rho_{b0}$); 
Bahcall et al. (1992) considered several models in which the dark matter was distributed like different components 
of matter in the disk
and found $\rho_0/\rho_{b0}\geq 2$ for the best fitting ones.

A high density of dark matter in the disk has been invoked to explain several terrestrial phenomena with periodicities of order 30 Myr.
To our knowledge, the first such proposal was that of
Rampino \& Stothers (1984), who  suggested that the then-estimated periodicity in mass extinctions of 30 Myr 
(Raup \& Sepkoski 1984) could be
naturally explained as the half-period of solar oscillations about the Galactic plane if $\rho_0\simeq 0.2\mpce, which was within
the range of observed values at that time; they further suggested that ``missing matter" might contribute to this density.
Shaviv et al. (2014) inferred $\rho_0=0.21\mpce\ in order to explain
the 32 Myr period observed in calcitic fossil shells; they attribute the excess density over the observed $\rho_0\sim 0.1\mpce\ 
to a disk dark-matter component
that is distinct from the standard halo dark matter. Fan et al. (2013) suggested that a more complex dark sector could include a small fraction of dark matter that is dissipative, so that it could cool and form a disk, like baryons do; they termed this ``Double Disk Dark Matter."
Randall \& Reece (2014) explored a particular version of this model in which 
the disk of dissipative dark matter is very thin, so that comet showers are induced when the Sun passes through the disk
with a half-period that matches the 35 Myr periodicity they
infer from the record of large impact craters on Earth. 
In fact, as we shall show, all these models require that the disk of dark matter be significantly thinner than the baryonic disk.
We define a ``thin" dark disk as one with a thickness $\la$ that of the stellar disk ($\sim 400$~pc, Eq. (\ref{eq:hstar})).
This dark disk is therefore distinct from the one conjectured to be produced by accreted satellite galaxies, 
which could create
a disk of dark matter with a density in the range $0.25-1.5$ times that of the halo dark matter
but with a thickness $\ga$ 1 kpc (Read et al. 2008, 2009).

We obtain an approximate
upper bound on
the surface density of a possible disk of dark matter 
by comparing the dynamically determined disk surface density with that determined from star counts and the observed amount
of gas (Kuijken \& Gilmore 1989b, 1991). The determinations of $\Sigma_b$ in the ``Stellar Dynamics" section of Table \ref{tab:baryon} are based on the assumption that
the matter within a distance $\sim$ 1 kpc of the Galactic plane is divided into a disk component, labeled ``baryons," and a dark
matter component, which has a constant density. The ``baryonic" component thus 
includes a possible disk of dark matter that has a scale height
similar to or less than that of the baryons;
we label the surface density of this component within a distance $z$ of the midplane $\Sigma_{\disk,\,z}$.
 Our value for $\Sigma_b$ is a valid measure of the baryons since it is based on (1) counts
of the M dwarfs, (2) a determination of the surface density of the remaining stars that is consistent with the PDMF of RGH02, and (3)
observations of the gas. Hence we can estimate the surface density of a possible disk of dark matter, $\Sigma_\ddm$, by subtracting
our determination of $\Sigma_b$ from the dynamical ones in Table \ref{tab:baryon},
$\Sigma_\ddm=\Sigma_{\disk,\,z}-\Sigma_{b,z}$. (Kramer \& Randall (2015) have shown that the presence of a thin dark disk can affect
the determination of the stellar surface density; this effect is small for small values of $\Sigma_\ddm$, and we ignore it for
this approximate discussion.)
 Before carrying out this procedure, 
 we note that the scale height of
the ionized gas, $h=1590$~pc, is so large that most of it would be counted as dark matter in a dynamical determination. Subtracting
the column density of HII from the values we measure, we obtain $\Sigma_{b,\,\rm disk}=44.4\mpc\ and $\Sigma_{b,\,\disk,1.1}=42.1\mpc.

The results of this exercise 
depend on the source of the data.
The results of Bienaym\'e et al. (2014) imply that the surface density of disk dark matter is $\Sigma_\ddm=0\pm 5.3\mpc\
(we have assumed that the disk surface density of $44.4\pm 4.1\mpc\ that they found is the total for the disk,
not the value for $z=1.1$~kpc).
At the other extreme,  the high value of the disk surface density found by Zhang et al. (2013), $55.2\mpc\ within
1.1 kpc, leads to $\Sigma_\ddm=13.1\pm 6.9\;\mpc. The average of all four estimates 
from the references in Table \ref{tab:baryon} is $\Sigma_\ddm=6.4\pm 6.8\mpc, which is consistent with both no thin disk of dark matter
and with an average 1 sigma upper bound  of $\Sigma_\ddm<13\mpc. 
This upper bound
is adequate to accommodate the model of Randall and Reece (2014), which has $\Sigma_\ddm\simeq (10-13)\mpc.

In order to determine if a thin dark disk is consistent with stellar kinematic data,
Kramer \& Randall (2015) have repeated the analysis of Holmberg \& Flynn (2000) with updated data and
an accounting for non-equilibrium effects. 
They point out that a thin disk of dark matter would gravitationally compress the stars, reducing the inferred surface density of stars with a given midplane density and velocity dispersion. Our determination of the stellar surface density is based on the midplane density and the observed scale height, so the sech$^2$ part of the distribution is reduced somewhat less than they infer and the exponential part is unaffected. Furthermore, our determination of the surface density of gas is independent of the possible existence of such a dark disk. An accurate determination of the effect of a thin dark disk on the inferred surface density of baryons is beyond the scope of this paper, however. Kramer \& Randall (2015) make the simplifying assumption that the baryons can be described by a single sech$^2$ distribution.
For an assumed scale height for the dissipative dark matter
of 20 pc (they adopt a definition for the
scale height that differs by a factor 2 from ours and give the scale height as 10 pc), they find a best-fit
surface density of $4\mpc\ for the dark disk, which is less than the value needed to account for
a terrestrial 30 Myr periodicity. They also find a surface density of visible stars of $33\pm 2 \;\mpc,
marginally consistent with the result of Bovy et al. (2012b) but higher than the value we found above, and a total surface density
within 1.1 kpc of $69\pm 6\;\mpc, in good agreement with the results in Table \ref{tab:dm}.

As noted above, the scale height of a disk of dark matter with a midplane density comparable to the baryon density is
significantly smaller than the baryonic scale height:
\beq
h_\ddm=\frac{\Sigma_\ddm}{2\rho_{0,\ddm}}=\frac{\Sigma_\ddm}{2(\rho_{0,\tot}-\rho_{b0}-\rho_{\rm halo~DM})},
\eeq
where $\rho_{0,\tot}$ includes the density of the disk dark matter
and $\rho_{\rm halo~DM}$, which is denoted as $\rho_\dm$ in the rest of the paper, is the midplane density of the halo dark matter.
In order to bring the half-period
of solar vertical oscillations down to about $P_{1/2}=30$~Myr, the total midplane density must be about $0.2\mpce\
(Rampino \& Stothers 1984) if the density is constant with height and greater than this if the
scale height is not large compared to the amplitude of the solar motion.
\footnote{If the Sun has a vertical velocity of 7.25 km s\e\
(Sch\"onrich et al. 2010) and is 14 pc above the midplane (Binney et al 1997), then the amplitude of the solar oscillation about the midplane is $72[P_{1/2}$/(30 Myr)] pc under the assumption that the motion is sinusoidal.}
For $\Sigma_\ddm<13\mpc\ this gives
$h_\ddm<63$ pc
if we use our result (Eq. \ref{eq:rhodm2}) for the density of the halo dark matter; in fact, the presence of disk dark matter would
reduce the inferred density of the halo dark matter and slightly reduce $h_\ddm$.
The model considered by Randall \& Reece (2014), which has $h_\ddm\sim (5-10)$~pc, readily satisfies this condition. 
A scale height $\la 60$~pc would require a midplane density considerably greater than $0.2\mpce\ in order to maintain a half-period
of about 30 Myr;
the numerical values of the Shaviv et al. (2014) model would require revision to accommodate the density variation in the dark
disk.

Before closing we note that while very thin dark disks are consistent with existing data, they face significant theoretical problems.
The first is that the dark matter must have unusual properties, as in the model of Fan et al. (2013), 
in order to be able to collapse into a disk that is significantly
thinner than the baryonic disk. Second, very thin sheets are cold and therefore gravitationally unstable
(Toomre 1964). Galactic disks adjust themselves so that
they are at least marginally stable against gravitational instabilities (e.g., Quirk 1972), so it is difficult for a self-gravitating
structure to be significantly thinner, and therefore colder, than the gas in the disk. Further work on the physics of the
proposed dark disks is required.

\section{Conclusions}

{\it Stars.} About half the stellar mass in the disk of the Milky Way
is in the form of M dwarfs. The column density of M dwarfs in the solar neighborhood was
determined by star counts in images taken with the Hubble Space Telescope (HST)
in a series of papers by Bahcall, Flynn, Gould and collaborators, the last of which was 
Zheng et al. (2001; Z01). We have revised the estimates of
the surface density in those papers by using the determination of the density of stars in the immediate solar neighborhood
(within 25 pc) by Reid et al. (2002; RGH02), obtaining an M-star surface density of $17.3\pm 2.3\, M_\odot$~pc\ee.

We have determined the surface densities of main sequence stars more massive than M dwarfs by using the local densities
determined by RGH02, the M-star scale height based on the results of Z01, and the relative scale heights obtained
by Flynn et al. (2006; F06). The overall normalization of the F06 scale heights appears too large since it disagrees with the values
obtained by Z01, \juric et al. (2008), Bochanski et al. (2010), Bovy et al. (2012b,c) and Bovy \& Rix (2013). 
We therefore reduced the F06 scale heights by the factor required
to bring the effective M-star scale height down to the Z01 value of 400 pc. This leads to a total surface density of visible stars
(main sequence stars and giants) of $27\pm 2.7\,M_\odot$~pc\ee. This is marginally consistent with the result
of Bovy et al. (2012b), who found $30\pm 1\,M_\odot$~pc\ee. However, the two results become
fully consistent if allowance is made for the fact that their error estimate
does not include the uncertainty in determining the mass range of the G stars they sampled.

We have calculated the density of white dwarfs under the assumptions that the IMF above $1\,M_\odot$ is
a power law, $\psi(m)d\ln m\propto m^{-\Gamma}d\ln m$ and that the star formation rate in the disk can be approximated
as a linear function of time. The white-dwarf density is relatively insensitive to the parameter describing this star formation history,
$b_0$, which is the ratio of the current rate to the average rate. For $b_0=0.5$, we find $n_{*,\WD}=8.5\times 10^{-3}$~pc\eee.
Katz et al. (2014) suggested that the density of white dwarfs has been significantly underestimated since the fraction of faint white
dwarf companions to main sequence stars is much less than the fraction of bright white dwarf companions. Based on their
discussion, we estimate that the density of white dwarfs is $8.3\times 10^{-3}$~pc\eee, very close to the value obtained from 
direct calculation and larger than many previous estimates---e.g., $6.0\times 10^{-3}$~pc\eee\ (Reid 2005)
and $4.9\times 10^{-3}$~pc\eee\ (Sion et al. 2009). Since on average the progenitors of white dwarfs are older than a typical
low-mass star, their scale height is somewhat larger (430 pc vs. 400 pc). For a mean white-dwarf mass of $0.665\,M_\odot$
(Holberg et al. 2008), the surface
density of white  dwarfs is $\Sigma_{*,\WD}=4.9\,M_\odot$~pc\ee.

Combining these estimates for the surface densities of main sequence stars, giants and white dwarfs with estimates
of the surface densities of brown dwarfs, neutron stars and black holes (Table \ref{tab:local}), we find a total stellar surface density 
$\Sigma_*=33.4\pm 3\,M_\odot$~pc\ee. This is close to the result of F06, who found $\Sigma_*=35.9\,M_\odot$~pc\ee.
Our results for visible stars are almost identical to theirs, although our mass distribution has significantly more M dwarfs
than theirs; their estimate of
the surface density of stellar remnants (mainly white dwarfs) is considerably larger than ours and therefore much
larger than those of other workers, and they also estimated a surface density of brown dwarfs that is almost twice our value.
Table \ref{tab:local} also gives the effective scale heights ($h=\Sigma/2\rho_0$) for the different stellar and gas components,
which can be useful in modeling the gravitational field normal to the plane in the solar vicinity.

{\it Gas.} The properties of the local interstellar medium are summarized in Table \ref{tab:ism}. The principal uncertainty
in the column density of interstellar gas has been the HI (Read 2014). We have carefully analyzed the existing observational
data on the local HI and concluded that the local surface density of optically thin HI is $8.3\,M_\odot$~pc\ee\ based on
the observations of Heiles (1976) and Heiles et al. (1981), including He and heavier elements. This is essentially the
same as the value of $8\,M_\odot$~pc\ee\ given by Kulkarni \& Heiles (1987) and is quite close to
the value of $9\,M_\odot$~pc\ee\ cited by Kalberla (private communication) based on the data in Kalberla \& Dedes (2008).
Based on the results of Heiles \& Troland (2003), Strasser \& Taylor (2004), and Braun (2012), we estimated that the
optical depth in the 21 cm line implies that the total amount of HI is 1.3 times the optically thin value; for the CNM,
which is the optically thick component of the HI, we inferred a correction factor of 1.68.
We then found that the
total surface density of local HI is 
$10.9\pm 1.6\,M_\odot$~pc\ee. This is significantly less than the value cited by Read (2014) of
$17\pm 4\,M_\odot$~pc\ee\ since he took the surface density of HI to be $12\,M_\odot$~pc\ee\ from Kalberla
\& Dedes (2008) and, based on that reference, then added the mass of the He and heavier elements. We constructed
a model for the distribution of local HI based on the work of Heiles and collaborators and showed that this model is
consistent with the results inferred from {\it Planck} data by Fukui et al. (2015). This agreement has the important 
implication that our accounting of the local interstellar gas is not missing a significant amount of dark gas.
Including the local surface density of \htwo\ and ionized hydrogen, we find a total gas surface density of 
$\Sigma_g=13.7\pm 1.6 \,M_\odot$~pc\ee,
in good agreement with the $13\mpc\ given by Bahcall et al. (1992). 
The uncertainty in this value is significantly less than that in Read's (2014) estimate.
It is often assumed that all this gas is close to the Galactic
plane, but the ionized hydrogen extends far from the plane (Gaensler et al. 2008, Schnitzler et al. 2012); we find that the
surface density of gas within 1.1 kpc of the plane is $12.6\mpc.

{\it Baryons.} Our results for the local surface densities of stars, gas and baryons are given in Table \ref{tab:baryon}, together with
the results of a number of other workers. Our result for the total baryon surface density, excluding halo stars at $z> 3$ kpc and
hot halo gas,
$\Sigma_b= 47.1\pm 3.4\mpc,
agrees with the results of the other references cited in the table with the exception
of Read (2014), whose estimate of the gas surface density is much larger than ours as explained above.
Our result for the baryon surface density within 1.1 kpc of the Galactic plane is significantly less than that of Zhang et al. (2012),
who found $\Sigma_b=55\pm 5\mpc. 

{\it Dark Matter.} The local mass density of dark matter can be determined by subtracting the surface density of baryons from
the total surface density of matter determined from an analysis of stellar kinematics. In Table \ref{tab:dm} we present the
results from several references together with the results that would be obtained with our value for the baryon surface density.
The average value we find from this exercise is $\rho_\dm=0.013\pm 0.003\mpce, which is consistent with the results of
Bienaym\'e et al. (2014) and Piffl et al. (2014), and with that of Bovy \& Tremaine (2012) after including their corrections
for the difference in the values of $\rho_\dm$ in the midplane and at a distance of 2.5 kpc from the plane. 
Our result is also consistent with the global measures of the dark matter density discussed by Read (2014).
We find that the total 
local density of matter is
$\rho_0=0.097\pm 0.013\mpce.
The different determinations of the surface density of disk dark matter that we have analyzed lead to inconclusive results
on the existence of a thin disk of dark matter.
Using the method of Kuijken \& Gilmore (1989b, 1991), we find that
the average 1 sigma upper bound on a possible thin disk of dark matter in the Galactic plane is $\Sigma_\ddm\la 13\mpc, which
is adequate to accommodate the models of Shaviv et al. (2014) and Randall \& Reece (2014). 
This implies that if the midplane density is at least $0.2\mpce, which is necessary in order to match the the periodicities of
several terrestrial phenomena, then the effective scale height of the disk dark matter is $\la 60$ pc. This is consistent with
the model of Randall \& Reece (2014) and of Kramer \& Randall (2015), but would require revision of the 
numerical parameters of the model of Shaviv et al. (2014).

{\it Determining the Surface Density $\Sigma$ from the Acceleration $K_z$.}
Dynamical determinations of the total surface density of matter generally begin by determining the vertical acceleration due to gravity
at some height $z$ above the plane, $K_z$.
Appendix C gives the relation between $\Sigma$ and $K_z$ and shows that it is in good agreement with the numerical results of
Bovy \& Rix (2013).

\acknowledgments

We wish to particularly thank Jo Bovy for a number of helpful comments on this paper, and Peter Kalberla, Carl Heile
and Yasuo Fukui
for clarifying the
HI observations.
We also 
thank Joss Bland-Hawthorn, Jim Cordes,
John Dickey, 
Chris Flynn, Andrew Gould, 
David Hogg, Chung-Pei Ma,
Lisa Randall,
Justin Read, Neil Reid, Nicola Sartore, Scott Tremaine, Mark Wolfire
and Zheng Zheng for helpful remarks.
AP was supported for part of this research as a Senior Associate by the National Research Council, as guest researcher by the IESA-CSIC, and by the CDCHTA of the Universidad de Los Andes.
The research of CFM is supported in part by NSF grant AST-1211729
and by NASA grant NNX13AB84G.
We acknowledge early support from the NASA Astrophysical Theory Program in
RTOP 344-04-10-02.

\appendix

\section{Effects of the Metallicity Gradient on Stellar Scale Heights}
\label{app:scale}

The mean stellar metallicity decreases with distance from the midplane, which affects both
the color-magnitude relation (CMR) used to estimate distances and the mass-luminosity relation
used to infer stellar masses.
Stars of a given luminosity are
less massive at lower metallicity. Recall that we define the transition
mass between M and K stars, $\mmk$, to be at $M_V=8$, as in Gould et al (1996). This transition mass
decreases with decreasing metallicity, dropping by about
17\% at the point at which the metallicity is 0.5 dex below solar (Baraffe et al. 1998).

Let $dn_*(m,z)$ be the number density of stars in the mass range $m$ to $m+dm$ at
a distance $z$ from the galactic plane; we denote the value at the plane as
$d\nnso(m)$.  For
stars above the hydrogen-burning limit ($m>m_\BD$),
we take $m$ to be the Zero Age Main Sequence (ZAMS) mass, measured in solar masses.
The number density of M dwarfs at height $z$ is
\beq
\nnsm(z)=\int_\mbd^{\mmk(z)} \frac{dn_*(m,z)}{d\ln m}\, d\ln m.
\eeq
The value at the midplane is $\nnsmo$, and the corresponding mass density of M dwarfs  
at the midplane is $\rhosmo$.
Let
\beq
\frac{dN_*(m)}{d\ln m}=\int_{-\zmax}^{\zmax} \frac{dn_*(m,z)}{d\ln m}\,dz
\eeq
be the number surface density of stars per unit logarithmic mass interval within a distance $|\zmax|$ of the midplane. 
We shall assume that the stars are distributed symmetrically about the midplane. The number surface density of M dwarfs is then
\beqa
N_{*,\M}&=& 2\int_0^\zmax \nnsm dz,\\
&=&2\int_0^\zmax dz\int_{\mbd}^{{\mmk(z)}} \frac{dn_*(m,z)}{d\ln m}\, d\ln m,
\label{eq:Nsm}
\eeqa
and the corresponding mass surface density is $\SSM$.
For stars with lifetimes greater than that of the disk, such as M dwarfs, we can express the number distribution in terms of the IMF 
(eq. \ref{eq:imf}) as
\beq
\frac{dn_*(m)}{d\ln m}=\left(\frac{\nnsmo}{F_\M}\right)\psi(m),
\label{eq:imf1}
\eeq
where
\beq
F_\M=\int_\mbd^\mmk \psi(m) d\ln m
\eeq
is the number fraction of M dwarfs in the IMF at the midplane and where $\mmk=\mmk(0)$.

The stellar metallicity gradient with distance from the plane implies that the
vertical structure of the disk can be characterized by three distinct effective scale heights:
\begin{itemize}
\item The effective scale height for a fixed mass, 
\beq
h(m)=\frac{\ssm}{2 d\nnso(m)/d\ln m}.
\eeq
\item The effective scale height for a fixed absolute magnitude; since we follow 
Zheng et al. (2001) in identifying
stars with $M_V\geq 8$ as M dwarfs and since all M dwarfs have the same scale height,  the M-dwarf scale height is
\beq
h_\M=\frac{N_{*,\M}}{2\nnsmo},
\label{eq:hm}
\eeq
where $N_{*,\M}$ is the number of M dwarfs per unit area identified by their absolute magnitudes.
\item The mass-weighted effective scale height,
\beq
h_{m,\M}=\frac{\SSM}{2\rhosmo},
\label{eq:hmma}
\eeq
where $\SSM$ is the mass surface density of M dwarfs.
\end{itemize}
One could also define an effective scale height based on the stellar colors, as done by \juric et al.
(2008), for example. For a color-magnitude relation that is independent of height, as they assumed,
this scale height corresponds to a fixed absolute magnitude and is similar to $h_\M$.
In the following we show that the scale heights defined above are within a few percent of one another,
so that the distinction among them, while important in principle, is not significant in practice.

Given our assumption of a universal IMF, the effect of a variation in metallicity on the scale heights
of M dwarfs is due to its effects on their mass range. The minimum mass of an M dwarf,
$\mbd$, is very weakly dependent on metallicity (Burrows et al. 2001), so we neglect this
variation. However, as noted above, the maximum mass of an M dwarf, $\mmk$, decreases by about
17\% as the metallicity decreases from solar by 0.5 dex (Baraffe et al. 1998). 
For simplicity, we represent the density distribution of low-mass stars with a single exponential,
\beq
dn_*(m,z)=dn_*(m,0)\exp(-z/h_0). 
\eeq
With the aid of Equation (\ref{eq:imf1}), Equation (\ref{eq:Nsm}) then becomes
\beq
N_{*,\M}=2\nnsmo h_0 -\frac{2\nnsmo}{F_\M}\int_0^\zmax e^{-z/h_0}dz \int_{\mmk(z)}^\mmk
\psi(m) d\ln m.
\eeq
Since the variation of $\mmk(z)$ is not large, we obtain
\beq
\frac{h_\M}{h_0}\simeq 1-\frac{\psi(\mmk)}{F_\M h_0}\int_0^\zmax e^{-z/h_0}\ln\left[\frac{\mmk}{\mmk(z)}\right]
dz.
\label{eq:hm1}
\eeq

Using the
results of Baraffe et al. (1998), Z01 modeled the metallicity gradient as a linear variation in the mass-luminosity relation between
solar metallicity at the midplane and a metallicity lower by 0.5 dex at $z=1500$~pc; they assumed a constant metallicity above 1500 pc. We implement
their model by taking
\beq
\ln\left[\frac{\mmk}{\mmk(z)}\right]=\frac{z}{1500\mbox{ pc}}\ln\left[\frac{\mmk}{\mmk(1500\mbox{ pc})}\right].
\label{eq:ln}
\eeq
We estimate that the logarithmic factor on the RHS is 0.16 from the results of Baraffe et al. (1998).
As discussed in the text, Z01 found an effective scale height for M dwarfs of $h_\M\simeq 400$~pc,
and we adopt that value for $h_0$. Since this is small compared with 1500 pc, we apply
Equation (\ref{eq:ln}) beyond 1500 pc and let $\zmax\rightarrow\infty$. Since $\mmk=0.67$,
Equation (\ref{eq:hm1}) becomes
\beq
\frac{h_\M}{h_0}\simeq 1-\frac{0.16\psi(0.67)}{F_\M}\left(\frac{h_0}{1500\mbox{ pc}}\right).
\eeq
For the standard PMH11 IMF, $\psi(0.67)=0.179$ and $F_\M=0.555$, so that for $h_0=400$~pc,
\beq
\frac{h_\M}{h_0}=0.986.
\eeq

One can repeat the same analysis for
\beq
\SSM=2\int_0^\zmax dz\int_{\mbd}^{{\mmk(z)}} m\,\frac{dn_*(m,z)}{d\ln m}\, d\ln m,
\eeq
and obtain 
$h_{m,\M}/h_0=\SSM/(2\rhosmo)=0.966$. The ratio of $h_{m,\M}$ to $h_\M$, 
the quantity measured by Z01, is 0.980. We conclude that these three different scale heights
are approximately the same in practice.

\section{The Stellar Halo}
\label{app:halo}

In their extensive study of the stellar number density distribution in the Milky Way based on data from the Sloan Digital
Sky Survey, \juric et al. (2008) found clear evidence for a stellar halo. In our notation, their representation of the
vertical stellar number-density distribution at the solar circle is
\beq
n_*=n_{*,0}\left\{(1-\beta)\exp(-z/h_1)+\beta\exp(-z/h_2)+(1-\beta)f_h\left[1+\left(\frac{z}{q_hR_0}\right)^2\right]^{-\nu}
\right\}.
\label{eq:haloden}
\eeq
The last term represents the stellar
halo; the radius of the solar circle is $R_0$, which they took to be 8 kpc, 
$f_h$ is proportional to the fraction of stars in the halo,
and the number $q_h$ measures the deviation from
sphericity of the halo. The corresponding stellar surface density is
\beq
N_*=2n_{*,0}\left[(1-\beta)h_1+\beta h_2+\frac 12 (1-\beta)f_h q_hR_0\rm{B}\left(\frac 12,\nu-\frac 12\right)\right],
\eeq
where B$(x,y)$ is the beta function.

\juric et al. (2008) presented two fits to their data. Redder stars ($r-i>1.0$) are too faint to have a significant halo contribution;
we refer to the fit obtained from these data 
as the ``disk fit," which has $f_h=0$ by assumption. The scale heights are $h_1=245$ pc and $h_2=743$ pc,
and the thick/thin disk ratio is $\beta/(1-\beta)=0.13$. The fit to the distribution of the bluer stars for $R<20$ kpc includes a halo
component; we refer to this as the ``disk + halo  fit." This fit has $f_h=0.0051$, $q_h=0.64$ (i.e., a flattened halo),
$\nu=1.4$, $h_1=251$ pc, $h_2=647$ pc and
$\beta/(1-\beta)=0.12$. Corrections for binarity and Malmquist bias were presented only for the disk fit, so these
values of the parameters do not include these corrections; it is not clear how these corrections
would affect the inferred halo parameters. We assume that these corrections have at most a modest
effect on the ratio of halo stars to disk stars. For example,
halo stars contribute 8\% of the total surface density in the disk + halo fit, 
and we anticipate that this result would be the
very nearly the same if the corrected values of the scale heights, etc., were used.

\juric et al. (2008) suggested that the disk fit includes many halo stars, and we have confirmed that: the surface
density for the two fits agrees to within 0.4\% for 1 kpc $\la z\la$ 3 kpc. At greater heights, the surface density of
the disk + halo fit grows steadily larger than that of the disk fit, eventually being 4.4\% larger. We conclude that
including all the halo stars increases the surface density by about 4\% above that obtained from the disk
fit. This ``high-halo" ($z>3$~kpc) contribution includes
about half the total number of halo stars.

We assume that the halo was formed $\sim 10$ Gyr ago, so that only stars with $M_V\ga 5$ survive. The surface
density of such stars in the disk is 23.1 $M_\odot$ pc\ee\ according to Table 1, so the surface density of main sequence stars
in the high halo is 4\% of this, or $0.92\, M_\odot$ pc\ee. Brown dwarfs contribute $0.05\, M_\odot$ pc\ee. 
The contribution of
white dwarfs is somewhat larger than 4\% since all stars with $8\ga m\ga1$ have become white dwarfs; the parameter
$a_1$ in Equation (\ref{eq:a1}) is 0.7 for a starburst 10 Gyr ago instead of 0.53 for the case we considered
in Section \ref{sec:wd}. We estimate a white-dwarf surface density of $0.35\, M_\odot$ pc\ee\ for the high halo. 
The surface density of neutron stars and black holes in the high halo is negligible, so we ignore that. Altogether,
we estimate that the high halo contributes a surface density of $1.3\, M_\odot$ pc\ee. The total halo contribution
is about $2.6\,M_\odot$ pc\ee, but half of that is included in the disk fits. 

Most studies of the stellar surface density focus on stars below 3 kpc and
assume that the density profile can be described by one 
(e.g., Zhang et al. 2013) or two (e.g.,  Bochanski et al. 2010)
exponentials, or the sum of an exponential and a $\sech^2$ function (e.g., Z01). Halo stars within the survey volume
are then automatically included in fits to the disk. We have therefore excluded halo stars above 3 kpc from
our analysis. By contrast, F06 explicitly included a halo component extending to the midplane. With
an effective scale height $h=\Sigma_*/2\rho_{*0}=3$~kpc, about 2/3 of these halo stars are within 3 kpc
of the disk, and these have been included in Table \ref{tab:local}.

Although not of direct relevance to this paper, it is of interest to know the total mass of the stellar halo implied
by the results of \juric et al. (2008).
Their general expression for the stellar density in the halo is given by Equation (\ref{eq:haloden}) with ``1" in
the final factor replaced by $(R/R_0)^2$, where $R$ is the cylindrical radius. 
The mass in the stellar halo can then be obtained by integrating this equation over an ellipsoidal volume.
\juric et al. (2008) state that their data extend out to galactocentric distances of about 20 kpc. Interpreting this
distance as the height, $Z$, of an ellipsoidal volume, we find the mass of the halo within that volume to be
\beqa
M_h&=&\left(\frac{4\pi}{3-2\nu}\right)q_h f_h 
\rho_{*,0}(1-\beta)R_0^{2\nu}\left(\frac{Z}{q_h}\right)^{3-2\nu},\\
&\rightarrow& 5.1\times 10^9\left(\frac{R_0}{8\mbox{ kpc}}\right)^{2.8}\left(\frac{Z}{20\mbox{ kpc}}\right)^{0.2}~~~M_\odot,
\eeqa
for the parameter values adopted by \juric et al. (2008). In the final expression, we have set the midplane density of disk stars to be
$0.0367\,M_\odot$~pc\eee, which includes all the stars and stellar remnants with lifetimes $>10^{10}$ yr from Table \ref{tab:local}.

\section{Relating the Total Surface Density, $\Sigma$, and the Gravitational Acceleration, $K_z$}
\label{app:kz}

Under the assumption of axisymmetry,
Poisson's equation, $\div\vecK=-4\pi G\rho$, in cylindrical coordinates $(R,z)$ is
\beq
\ppbyp{K_z}{z} - \frac{1}{R}\pbyp{R}v_c^2=-4\pi G\rho,
\eeq
where $v_c^2=-RK_R$ is the circular velocity at $R$. Integration from $-z$ to $z$ (assumed positive) gives 
\beq
K_z-\frac{1}{R}\pbyp{R}\int_0^z v_c^2 dz=-2\pi G\Sigma
\label{eq:sig}
\eeq
(cf. Kuijken \& Gilmore 1989a, who made the approximation that $v_c$ is independent of $z$).
Defining
\beq
\Delta\Sigma\equiv \frac{1}{2\pi GR}\pbyp{R}\int_0^z v_c^2 dz',
\label{eq:deltas}
\eeq
this becomes
\beq
\Sigma=-\frac{K_z}{2\pi G}+\Delta\Sigma.
\label{eq:kz}
\eeq

It is often stated that if the rotation curve is flat, then $\Delta\Sigma=0$ so that $|K_z|=2\pi G\Sigma$.
However, this is not correct, since if the rotation curve is flat in the midplane, it is not flat away from the 
plane. Following Bovy \& Tremaine (2012), we note that
\beq
\ppbyp{K_R}{z}=-\pbyp{z}\ppbyp{\phi}{R}=-\pbyp{R}\ppbyp{\phi}{z}=\ppbyp{K_z}{R},
\eeq
where $\phi$ is the gravitational potential. Replacing $K_R$ with $-v_c^2/R$ and using Equation (\ref{eq:kz}),
we obtain
\beq
v_c^2=v_{c0}^2+2\pi G R\pbyp{R}\int_0^z(\Sigma-\Delta\Sigma)dz',
\eeq
where $v_{c0}=v_c(z=0)$ is the circular velocity in the midplane. Inserting this into Equation (\ref{eq:deltas}) gives
\beq
\Delta\Sigma=\frac{1}{2\pi GR}\pbyp{R}\int_0^z dz'\left[v_{c0}^2+2\pi G R\pbyp{R}\int_0^{z'}(\Sigma-\Delta\Sigma)dz''\right].
\label{eq:deltasv}
\eeq
Generally, only the first term is retained (e.g., Kuijken \& Gilmore 1989a).

Consider each of these terms in turn. The first term, labeled $\Delta\Sigma_1$, is
\beq
\Delta\Sigma_1=\frac{\alpha zv_{c0}^2}{\pi G R^2},
\label{eq:deltas1}
\eeq
where 
\beq
\alpha\equiv\ppbyp{\ln v_{c0}}{\ln R}.
\eeq
In their analysis of the Milky Way's circular-velocity
curve between 4 and 14 kpc, Bovy et al. (2012a) find
\beq
\alpha=0.01^{+0.01}_{-0.1}.
\label{eq:alpha}
\eeq
For a flat rotation curve, $\alpha$ and therefore $\Delta\Sigma_1$ vanish, but the remaining two terms in Equation (\ref{eq:deltasv})
do not, so that $|K_z|\neq 2\pi G\Sigma$. Numerically, $\Delta\Sigma_1$ is less than 10\% of the local surface density, $\Sigma\simeq 70\mpc,
for $z\la 1.1$~kpc:
\beq
\Delta\Sigma_1=5.72\left(\frac{\alpha_{0.1}z_{1.1}v_{c0,220}^2}{R_{8.3}^2}\right)~~M_\odot\mbox{ pc\ee},
\eeq
where $\alpha_{0.1}\equiv \alpha/0.1$, $z_{1.1}\equiv z/(1.1$~kpc), $v_{c0,220}\equiv v_{c0}/(220$~km~s\e), etc. 
As $\alpha$ varies over the range allowed by observation,
$-5.15\mpc$\;<\Delta\Sigma_1<1.14\mpc\ 
for nominal values of the other parameters.
This term is often expressed in terms of the Oort constants $A$ and $B$ (Kuijken \& Gilmore 1989a),
\beq
\Delta\Sigma_1=-\frac{(A^2-B^2)z}{\pi G}.
\eeq

Next, consider the second term in Equation (\ref{eq:deltasv}),
\beq
\Delta\Sigma_2=\left(\frac{1}{R}\pbyp{R}+\frac{\partial^2}{\partial R^2}\right)\int_0^z dz'\int_0^{z'}dz''\,\Sigma.
\label{eq:dsig21}
\eeq
We adopt a double-exponential form for the density, including the dark matter, in the solar neighborhood:
\beq
\rho=\rho_{00}e^{(-R/h_R-z/h_z)},
\eeq
with $h_z$ constant; note that $\rho_0=\rho_{00}\exp(-R/h_R)$.
The surface density is then
\beq
\Sigma=\Sigma_{00}e^{-R/h_R}\left(1-e^{{-z/h_z}}\right)=\Sigma_0\left(1-e^{{-z/h_z}}\right),
\eeq
where $\Sigma_0=2\rho_0 h_z$ is the total surface density at $R$. 
Since the dark matter is not exponential, this definition has to be modified in practice. 
We are interested in applying this equation in the solar neighborhood at a distance $\sim 1$~kpc from the plane, which
includes most of the baryons and where the surface density is dominated by the baryons. We therefore approximate $\Sigma_0$ as
the total surface density within a distance $z\simeq 1-1.1$ kpc of the plane, and in our numerical evaluations we set $\Sigma_0=70\mpc. 
With $\rho_0=0.093\mpce\ (Eq. \ref{eq:rhoo}), the effective scale height is $h_z=376$~pc.
Equation (\ref{eq:dsig21}) then becomes
\beq
\Delta\Sigma_2=\frac{z^2}{2h_R^2}\left(1-\frac{h_R}{R}\right)\left[1-\frac{2h_z}{z}+\frac{2h_z^2}{z^2}\left(1-e^{-z/h_z}\right)\right]\Sigma_0.
\label{eq:dsig22}
\eeq
Note that $\Delta\Sigma_2\geq 0$ for the cases of greatest interest, in which $R>h_R$ and $z>h_z$.
For nominal values of the parameters ($z=1.1$~kpc, $h_R=2.5$~kpc, $R=8.3$~kpc and $\Sigma_0=70 \mpc), we find
$\Delta\Sigma_2=2.54 \mpc. 
(We have found that applying the double-exponential model only to the baryons and using
an NFW model for the dark matter changes this result by only about 5\%.)
Hence, we have 
\beq
-2.6\;M_\odot\mbox{ pc\ee}<(\Delta\Sigma_1+\Delta\Sigma_2)<3.7 \;M_\odot\mbox{ pc\ee}
\label{eq:range}
\eeq
after allowing for the observed range of $\alpha$ (Eq. \ref{eq:alpha}).

Finally, we show that the third term in Equation (\ref{eq:deltasv}), 
\beq
\Delta\Sigma_3=-\left(\frac{1}{R}\pbyp{R}+\frac{\partial^2}{\partial R^2}\right)\int_0^z dz'\int_0^{z'}dz''\,\Delta\Sigma,
\label{eq:dsig3}
\eeq
is negligible. This is to be expected, since a rough self-consistency argument shows that $\Delta\Sigma_3$ is second order in $\Delta\Sigma/\Sigma$:
\beq
\Delta\Sigma_3\sim\left(\frac{\Delta\Sigma}{\Sigma}\right)\Delta\Sigma_2\sim \left(\frac{\Delta\Sigma}{\Sigma}\right)^2\Sigma,
\eeq
where the first step follows from Equation (\ref{eq:deltasv}) and the second step follows from the fact that $\Delta\Sigma_2$ is of order
$\Delta\Sigma$.
Because $\Delta\Sigma= \Delta\Sigma_1+ \Delta\Sigma_2+ \Delta\Sigma_3$, Equation (\ref{eq:dsig3}) becomes
\beq
\Delta\Sigma_3= \Delta\Sigma_{31}+ \Delta\Sigma_{32}+ \Delta\Sigma_{33}
\label{eq:dsig32}
\eeq
in an obvious notation. Since the rotation curve is approximately flat, we neglect the radial variation of the velocity and find
\beq
\Delta\Sigma_{31}\simeq -\frac 23 (1-\alpha)^2\left(\frac{z^2}{R^2}\right)\Delta\Sigma_1\simeq -0.012\Delta\Sigma_1,
\eeq
which is in the range $-0.014\mpc$<\Delta\Sigma_{31}<0.06\mpc\ for nominal values of the parameters and is negligible.
An upper limit on $\Delta\Sigma_2$ is
\beq
\Delta\Sigma_2<\left(\frac{z^2+2h_z^2}{2h_R^2}\right)\Sigma_0
\eeq
from Equation (\ref{eq:dsig22}). We then obtain
\beq
|\Delta\Sigma_{32}|<\left(\frac{z^4+12h_z^2z^2}{24 h_R^4}\right)\Sigma_0.
\eeq
For nominal values of the parameters, this upper bound is $0.26\mpc, which again is negligible.

To evaluate $\Delta\Sigma_{33}$, we must specify the radial and vertical dependence of the unknown term $\Delta\Sigma_3$.
We expect the radial dependence $\Delta\Sigma_3$ to be qualitatively similar to that of $\Delta\Sigma_1$, which varies
on the scale of the half-radius, $R/2$ (since $\Delta\Sigma_1\propto R^{-2}$), and
$\Delta\Sigma_2$, which varies on the radial scale length, $h_R$. Since $h_R\simeq 2.5\mbox{ kpc}\;<R/2\simeq 4$~kpc, we obtain
an approximate upper bound on $\Delta\Sigma_3$ by evaluating the radial derivatives with the radial scale length. Since
$\Delta\Sigma$ vanishes for $z=0$ according to Equation (\ref{eq:deltas}), we assume that $\Delta\Sigma_3\propto z^p$
for some value of $p\geq 0$. Equation (\ref{eq:dsig3}) then implies 
\beq
\Delta\Sigma_{33}\simeq -\frac{z^2\Delta\Sigma_3}{(p+1)(p+2)h_R^2}.
\eeq
Since $\Delta\Sigma_{33}$ has opposite sign to $\Delta\Sigma_3$, Equation (\ref{eq:dsig32}) implies that
\beq
|\Delta\Sigma_3|\la|\Delta\Sigma_{31}+\Delta\Sigma_{32}|.
\eeq
For nominal values of the parameters, we find $|\Delta\Sigma_3|\la 0.27\mpc, which is negligible.

We conclude that
\beq
\Sigma\simeq \frac{|K_z|}{2\pi G}+\Delta\Sigma_1+\Delta\Sigma_2,
\label{eq:kz2}
\eeq
where $-2.6\;M_\odot\mbox{ pc\ee}<(\Delta\Sigma_1+\Delta\Sigma_2)<3.7 \;M_\odot\mbox{ pc\ee}$ (Eq. \ref{eq:range})
for fiducial values of the parameters
as $\alpha$ varies over the allowed range (Eq. \ref{eq:alpha}).

We can test our result by comparing with the results of Bovy \& Rix (2013). 
We begin with their fiducial case, in which the rotation curve is flat 
locally
($\alpha=0$).
They give the values of $\Sigma$ and $K_z$ for a number of mono-abundance populations
measured at different Galactocentric radii. For the bin with data that is closest to the Sun
(7.5 kpc$\; < R <8$~kpc, where they adopted $R_0=8$~kpc), the average value of
$\Sigma-|K_z|/(2\pi G)$ is $2.6\mpc. Using 
their values of $h_z$ (400 pc) and $h_R$ (2.5 kpc, which we have adopted as our fiducial value) and using
the average value of $\Sigma_0$ in this bin
($73.4\mpc), we find $\Delta\Sigma\simeq \Delta\Sigma_2=2.5\mpc\ from Equation (\ref{eq:dsig22}),
in excellent agreement with their result. 
In making this comparison, we have ignored the small correction due to the fact that the slope of the rotation curve 
varies with radius in the Bovy \& Rix (2013) model (Jo Bovy, private communication).

They also considered the effects of non-flat rotation curves with $\alpha=\pm 0.1$
at the solar circle.
They adopted $v_{c0}=230$~km~s\e, so that
$\Delta\Sigma_1=\pm 6.7\mpc\ for $\alpha=\pm 0.1$.
For $\alpha=-0.1$, their analytic fits give $h_R=2.3$~kpc and
$\Sigma=65\,\mpc\ and $|K_z|/(2\pi G)=69\,\mpc, 
which implies $\Delta\Sigma_2=2.75\mpc. Given that $\Sigma=65\mpc,
we obtain $|K_z|/(2\pi G)=69.0\mpc, in excellent agreement with their result.
For $\alpha=+0.1$ they found $h_R= 2.7$~kpc and
$\Sigma=72\,\mpc, $|K_z|/(2\pi G)=63\,\mpc.
(The difference between $|K_z|/(2\pi G)$ and $\Sigma$ is larger than given in Equation (\ref{eq:range})
primarily since +0.1 is outside the observed range of $\alpha$ given by Bovy et al. (2012a).)
We then obtain $\Delta\Sigma_2=2.06\mpc; for $\Sigma=72\mpc, our
results give $|K_z|/(2\pi G)=63.2\mpc, which again is in excellent agreement with their result.

\end{document}